\documentclass[amssymb,11pt]{article}
\usepackage{amsmath}
\usepackage{amssymb}
\usepackage{graphicx}
\newtheorem{thm}{Theorem}[section]
\newtheorem{cor}{Corollary}[section]
\newtheorem{lem}{Lemma}[section]

\newtheorem{defn}{Definition}[section]
\newtheorem{prop}{Proposition}[section]
\newtheorem{exam}{Example}[section]

\def\>{\ensuremath{\rangle}}
\def\<{\ensuremath{\langle}}

\begin{document}

\title{Hoare Logic for Quantum Programs\thanks{This work was partly supported the National Natural Science
Foundation of China (Grant No: 60736011, 60621062) and the National
Key Project for Fundamental Research of China (Grant No:
2007CB807901)}}

\author{Mingsheng Ying\\University of Technology, Sydney and Tsinghua University\thanks{Author's address: Mingsheng Ying, Center of Quantum Computation and
Intelligent Systems, Faculty of Engineering and Information
Technology, University of Technology, Sydney, City Campus, 15
Broadway, Ultimo, NSW 2007, Australia, and State Key Laboratory of
Intelligent Technology and Systems, Tsinghua National Laboratory for
Information Science and Technology, Department of Computer Science
and Technology, Tsinghua University, Beijing 100084, China, email:
mying@it.uts.edu.au or yingmsh@tsinghua.edu.cn}}

\maketitle

\begin{abstract}
Hoare logic is a foundation of axiomatic semantics of classical
programs and it provides effective proof techniques for reasoning
about correctness of classical programs. To offer similar techniques
for quantum program verification and to build a logical foundation
of programming methodology for quantum computers, we develop a
full-fledged Hoare logic for both partial and total correctness of
quantum programs. It is proved that this logic is (relatively)
complete by exploiting the power of weakest preconditions and
weakest liberal preconditions for quantum programs.

\smallskip\

\textit{Keywords}: Quantum computation, programming language,
axiomatic semantics, Hoare logic, completeness
\end{abstract}

\section{Introduction}

Even though quantum hardware is still in its infancy, people widely
believe that building a large-scale and functional quantum computer
is merely a matter of time and concentrated effort. The history of
classical computing arouses that once quantum computers come into
being, quantum programming languages and quantum software
development techniques will play a key role in exploiting the power
of quantum computers. With expectation of offering effective
programming techniques for quantum computers, several quantum
programming languages have already been designed in recent years.
The earliest proposal for quantum programming language was made by
Knill \cite{Kn96}. The first real quantum programming language, QCL,
was proposed by $\ddot{O}$mer \cite{O03}; he also implemented a
simulator for this language. A quantum programming language in the
style of Dijkstra's guarded-command language, qGCL, was presented by
Sanders and Zuliani \cite{SZ00}. A quantum extension of C++ was
proposed by Bettelli et al \cite{BCS03}, and it was implemented in
the form of a C++ library. The first and very influential quantum
language of the functional programming paradigm, QFC, was defined by
Selinger \cite{Se04} based on the idea of classical control and
quantum data. In~\cite{AG05}, Altenkirch and Grattage defined
another functional programming language for quantum computing, QML,
in which both control and data may be quantum. For excellent survey
of quantum programming languages, see~\cite{Se04a,Ga06}.

The fact that human intuition is much better adapted to the
classical world than the quantum world is one of the major reasons
that it is difficult to find efficient quantum algorithms. It also
implies that programmers will commit much more faults in designing
programs for quantum computers than programming classical computers.
Thus, it is even more critical than in classical computing to give
clear and formal semantics to quantum programming languages and to
provide formal methods for reasoning about quantum programs. Indeed,
various semantic approaches to quantum programs have been proposed
in recent literatures. For example, an operational semantics was
given to Sanders and Zuliani's language qGCL~\cite{SZ00} by treating
an observation (quantum measurement) procedure as a probabilistic
choice; a denotational semantics was defined for Selinger's language
QPL~\cite{Se04} by interpreting quantum programs as super-operators;
and a denotational semantics of Altenkirch and Grattage's language
QML~\cite{AG05} was described in category-theoretic terms. In
addition, a language-independent approach to semantics of quantum
programs was proposed by D'Hondt and Panangaden~\cite{DP06} who
introduced an intrinsic notion of quantum weakest precondition and
established a beautiful Stone-type duality between state transition
semantics and predicate transformer semantics for quantum programs.

As to proof systems for reasoning about quantum programs, Baltag and
Smets~\cite{BS04,BS06,Ak05} presented a dynamic logic formalism of
information flows in quantum systems, which is capable of describing
various quantum operations such as unitary evolutions and quantum
measurements, and particularly entanglements in multi-partite
quantum systems. Brunet and Jorrand~\cite{BJ04} introduced a way of
applying Birkhoff and von Neumann's quantum logic~\cite{BvN36} to
the study of quantum programs by expanding the usual propositional
languages with new primitives representing unitary transformations
and quantum measurements. In~\cite{CMS}, Chadha, Mateus and Sernadas
proposed a Hoare-style proof system for reasoning about imperative
quantum programs using a quantitative state logic, but only bounded
iterations are allowed in their programming language. Feng et
al~\cite{FDJY07} found some useful proof rules for reasoning about
quantum loops, generalizing some of Morgan's proof rules for
probabilistic loops~\cite{Mor95}. To the author's best knowledge,
however, no complete Hoare logic for quantum programs has been
reported in the literature.

The main contribution of the present paper is the establishment of a
full-fledged Hoare logic for deterministic quantum programs based on
Selinger's idea of modeling quantum programs as super-operators and
D'Hondt and Panangaden's notion of quantum predicate as an Hermitian
operator~\cite{DP06}. This logic includes a proof system for partial
correctness and a proof system for total correctness of
deterministic quantum programs. In particular, we are able to prove
its (relative) completeness by exploiting the power of weakest
preconditions and weakest liberal preconditions for quantum
programs.

The paper is organized as follows: For convenience of the reader, we
recall some basic concepts of Hilbert spaces as well as the
fundamental postulates of quantum mechanics in Section 2. Another
aim of Section 2 is to fix notation used in the sequel sections. In
Section 3, we define the syntax of deterministic quantum programs
about which the Hoare logic presented in this paper is designed to
reason. Such quantum programs are quantum extension of classical
\textbf{while}-programs~(cf. \cite{AO97}, Chapter 3). In Section 4,
an operational semantics of quantum programs is given in terms of
transitions between quantum configurations, which consist of a
quantum program still to be executed and a (partial) density
operator expressing the current state of program variables. In
Section 5, we are able to introduce a denotational semantics of
quantum programs based on the operational semantics. A denotational
semantics of a quantum program is defined to be a function from
partial density operators to themselves. In Section 6, we adopt
D'Hondt and Panangaden's definition of quantum predicates as
Hermitian operators. Then a correctness formula is defined to be a
quantum extension of Hoare triple, which consists of two quantum
predicates, namely precondition and postcondition, as well as a
quantum program. Furthermore, the notions of partial and total
correctness can be introduced for quantum programs using their
denotational semantics. In Section 7, weakest precondition and
weakest liberal precondition for quantum programs are defined in
terms of total and partial correctness, respectively, in a familiar
way. With the long preparation of the previous sections, the Hoare
logic for quantum programs is finally established in Sections 8 and
9. In Section 8, a proof system for partial correctness of quantum
programs is presented, and its (relative) completeness is proved,
and in Section 9, after introducing the notion of bound function for
quantum loops, a proof system for total correctness of quantum
programs is given, and its (relative) completeness is also proved. A
brief conclusion is drawn and some open problems for further studies
are pointed out in Section 10.

\section{Preliminaries}

\subsection{Hilbert Spaces}

We write $\mathbb{C}$ for the set of complex numbers. For each
complex number $\lambda\in \mathbb{C}$, $\lambda^{\ast}$ stands for
the conjugate of $\lambda$. A (complex) vector space is a nonempty
set $\mathcal{H}$ together with two operations: vector addition $+:
\mathcal{H}\times \mathcal{H}\rightarrow \mathcal{H}$ and scalar
multiplication $\cdot : \mathbb{C}\times \mathcal{H}\rightarrow
\mathcal{H}$, satisfying the following conditions:
\begin{enumerate}\item $(\mathcal{H},+)$ is an abelian group, its zero element $0$ is called the zero vector; \item
$1|\varphi\rangle=|\varphi\rangle$; \item $\lambda (\mu
|\varphi\rangle)=\lambda\mu |\varphi\rangle$; \item $(\lambda
+\mu)|\varphi\rangle =\lambda |\varphi\rangle +\mu |\varphi\rangle$;
and \item $\lambda (|\varphi\rangle +|\psi\rangle)=\lambda
|\varphi\rangle + \lambda |\psi\rangle$ \end{enumerate} for any
$\lambda,\mu\in\mathbb{C}$ and $|\varphi\rangle,
|\psi\rangle\in\mathcal{H}$.

An inner product over a vector space $\mathcal{H}$ is a mapping
$\langle\cdot|\cdot\rangle:\mathcal{H}\times \mathcal{H}\rightarrow
\mathbb{C}$ satisfying the following properties:
\begin{enumerate}\item $\langle\varphi|\varphi\rangle\geq 0$ with
equality if and only if $|\varphi\rangle =0$; \item
$\langle\varphi|\psi\rangle=\langle\psi|\varphi\rangle^{\ast}$; and
\item $\langle\varphi|\lambda_1\psi_1+\lambda_2\psi_2\rangle=
\lambda_1\langle\varphi|\psi_1\rangle+\lambda_2\langle\varphi|\psi_2\rangle$\end{enumerate}
for any $|\varphi\rangle, |\psi\rangle, |\psi_1\rangle,
|\psi_2\rangle \in \mathcal{H}$ and for any $\lambda_1,\lambda_2\in
\mathbb{C}$. Sometimes, we also write
$(|\varphi\rangle,|\psi\rangle)$ for the inner product
$\langle\varphi|\psi\rangle$ of $|\varphi\rangle$ and
$|\psi\rangle$.

For any vector $|\psi\rangle$ in $\mathcal{H}$, its length
$||\psi||$ is defined to be $\sqrt{\langle\psi|\psi\rangle}$. A
vector $|\psi\rangle$ is called a unit vector if $||\psi||=1$. Let
$\{|\psi_n\rangle\}$ be a sequence of vectors in $\mathcal{H}$ and
$|\psi\rangle\in \mathcal{H}$. If for any $\epsilon
>0$, there exists a positive integer $N$ such that $||\psi_m
-\psi_n||<\epsilon$ for all $m,n\geq N$, then $\{|\psi_n\rangle\}$
is called a Cauchy sequence. If for any $\epsilon
>0$, there exists a positive integer $N$ such that $||\psi_n
-\psi||<\epsilon$ for all $n\geq N$, then $|\psi\rangle$ is called a
limit of $\{|\psi_n\rangle\}$ and we write $|\psi\rangle
=\lim_{n\rightarrow\infty}|\psi_n\rangle.$

A family $\{|\psi_i\rangle\}_{i\in I}$ of vectors in $\mathcal{H}$
is said to be summable with the sum $|\psi\rangle$ and we write
$|\psi\rangle=\sum_{i\in I}|\psi_i\rangle$ if for any $\epsilon>0$
there is a finite subset $J$ of $I$ such that
$$||\psi-\sum_{i\in K}\psi_i||<\epsilon$$ for every finite subset
$K$ of $I$ containing $J$. A family $\{|\psi_i\rangle\}_{i\in I}$ of
unit vectors is called an orthonormal basis of $\mathcal{H}$ if
\begin{enumerate}\item $|\psi_i\rangle\perp |\psi_j\rangle$ for any $i,j\in I$ with $i\neq j$; and \item
$|\psi\rangle=\sum_{i\in I}\langle\psi_i|\psi\rangle|\psi_i\rangle$
for each $|\psi\rangle\in\mathcal{H}.$\end{enumerate} In this case,
the cardinality of $I$ is called the dimension of $\mathcal{H}$.

A Hilbert space is defined to be a complete inner product space;
that is, an inner product space in which each Cauchy sequence of
vectors has a limit. According to a basic postulate of quantum
mechanics, the state space of an isolated quantum system is
represented by a Hilbert space, and a pure state of the system is
described by a unit vector in its state space.

\begin{exam}\label{ex-qub}\begin{enumerate}\item The state space of qubits is the
$2-$dimensional Hilbert space:
$$\mathcal{H}_2=\{\alpha|0\rangle+\beta|1\rangle:\alpha,\beta\in\mathbb{C}\}.$$
The inner product in $\mathcal{H}_2$ is defined by $$(\alpha
|0\rangle+\beta |1\rangle,\alpha^{\prime} |0\rangle+\beta^{\prime}
|1\rangle=\alpha^{\ast}\alpha^{\prime}+\beta^{\ast}\beta^{\prime}
$$ for all
$\alpha,\alpha^{\prime},\beta,\beta^{\prime}\in\mathbb{C}$. Then
$\{|0\rangle, |1\rangle\}$ is an orthonormal basis of
$\mathcal{H}_2$, called the computational basis.

\item The space $l_2$ of square summable sequences is $$\mathcal{H}_\infty=\{\sum_{n=-\infty}^{\infty}\alpha_n|n\rangle:\alpha_n\in\mathbb{C}\
{\rm for\ all}\ n\in\mathbb{Z}\ {\rm and}\
\sum_{n=-\infty}^{\infty}|\alpha_n|^{2}<\infty\},$$ where
$\mathbb{Z}$ is the set of integers. The inner product in
$\mathcal{H}_\infty$ is defined by
$$(\sum_{n=-\infty}^{\infty}\alpha_n|n\rangle,\sum_{n=-\infty}^{\infty}\alpha^{\prime}|n\rangle)=\sum_{n=-\infty}^{\infty}
\alpha_n^{\ast}\alpha_n^{\prime}$$ for all
$\alpha_n,\alpha_n^{\prime}\in\mathbb{C}$, $-\infty <n<\infty$. Then
$\{|n\rangle: n\in\mathbb{Z}\}$ is an orthonormal basis of
$\mathcal{H}_\infty$, called the computational basis.
\end{enumerate}\end{exam}

A (linear) operator on a Hilbert space $\mathcal{H}$ is a mapping
$A:\mathcal{H}\rightarrow\mathcal{H}$ satisfying the following
conditions:\begin{enumerate}\item
$A(|\varphi\rangle+|\psi\rangle)=A|\varphi\rangle+A|\psi\rangle$;
\item $A(\lambda |\psi\rangle)=\lambda A|\psi\rangle$
\end{enumerate} for all $|\varphi\rangle,|\psi\in\mathcal{H}$ and
$\lambda\in\mathbb{C}$. If $\{|\psi_i\rangle\}$ is an orthonormal
basis of $\mathcal{H}$, then an operator $A$ is uniquely determined
by the images $\{A|\psi_i\rangle\}$ of basis vectors
$\{|\psi_i\rangle\}$ under $A$. In particular, $A$ can be
represented by matrix
$$A=\left(\langle\psi_i|A|\psi_j\rangle\right)_{ij}$$ when
$\mathcal{H}$ is finite-dimensional.

An operator $A$ on $\mathcal{H}$ is said to be bounded if there is a
constant $C\geq 0$ such that $\|A|\psi\rangle\|\leq C\cdot\|\psi\|$
for all $|\psi\rangle\in\mathcal{H}$. In this paper, we only
consider bounded operators. We write $\mathcal{L}(\mathcal{H})$ for
the set of bounded operators on $\mathcal{H}$. The identity operator
on $\mathcal{H}$ is denoted $I_{\mathcal{H}}$, and the zero operator
on $\mathcal{H}$ that maps every vector in $\mathcal{H}$ to the zero
vector is denoted $0_\mathcal{H}$.

For any operator $A$ on $\mathcal{H}$, there exists a unique linear
operator $A^{\dag}$ on $\mathcal{H}$ such that
$$(|\varphi\rangle,
A|\psi\rangle)=(A^{\dag}|\psi\rangle,|\varphi\rangle)$$ for all
$|\varphi\rangle, |\psi\rangle\in\mathcal{H}$. The operator
$A^{\dag}$ is called the adjoint of $A$. An operator $M$ on
$\mathcal{H}$ is said to be Hermitian if $M^{\dag}=M$.

An operator $A$ on $\mathcal{H}$ is said to be positive if $\langle
\psi|A|\psi\rangle\geq 0$ for all states $|\psi\rangle\in
\mathcal{H}$. We can define a partial order between operators,
called the L$\ddot{o}$wner partial order : for any
$A,B\in\mathcal{L}(\mathcal{H})$, $A\sqsubseteq B$ if $B-A$ is a
positive operator.

\begin{lem}\label{order}$A\sqsubseteq B$ if and only if $tr(A\rho)\leq
tr(B\rho)$ for all density operators $\rho$.
\end{lem}

An operator $A$ is said to be a trace operator if
$\{\langle\psi_i|A|\psi_i\rangle\}_{i\in I}$ is summable for any
orthonormal basis $\{|\psi_i\rangle\}_{i\in I}$ of $\mathcal{H}$; in
this case, the trace $tr(A)$ of $A$ is defined to be
$$tr(A)=\sum_{i}\langle \psi_i|A|\psi_i\rangle$$ where
$\{|\psi_i\rangle\}$ is an orthonormal basis of $\mathcal{H}$. It
can be shown that $tr(A)$ is independent of the choice of
$\{|\psi_i\rangle\}$.

A density operator $\rho$ on a Hilbert space $\mathcal{H}$ is
defined to be a positive operator with $tr(\rho)=1$. Then a mixed
state of a quantum system with state space $\mathcal{H}$ is
described by a density operator on $\mathcal{H}$. In this paper, we
take a slightly generalized notion of density operator in the
sequel: a partial density operator $\rho$ is a positive with
$tr(\rho)\leq 1$. In particular, the zero operator is a partial
density operator. The set of partial density operators is denoted
$\mathcal{D}^{-}(\mathcal{H})$. A partial density operator can also
be defined by an ensemble of pure states. Suppose that a quantum
system is in one of a number of pure states $|\psi_i\rangle$, with
respective probabilities $p_i$, where it is required that
$\sum_ip_i\leq 1$. Then
$$\rho=\sum_ip_i|\psi_i\rangle\langle\psi_i|$$ is a density operator.
Conversely, any density operator can be generated in such a way.

\subsection{Unitary Transformations}

An operator $U$ on $\mathcal{H}$ is called a unitary transformation
if $U^{\dag}U=Id_{\mathcal{H}}$, where $Id_{\mathcal{H}}$ is the
identity operator on $\mathcal{H}$; that is,
$Id_{\mathcal{H}}|\psi\rangle=|\psi\rangle$ for all
$|\psi\rangle\in\mathcal{H}$.

\begin{exam}\begin{enumerate}\item The most frequently used unitary operators on qubits are the Hadamard transformation:
$$H=\frac{1}{\sqrt{2}}\left(\begin{array}{cc}1 & 1\\
1 & -1\end{array}\right),$$ and the Pauli matrices:
$$I=\left(\begin{array}{cc}1 & 0\\ 0 & 1\end{array}\right),\hspace{2em} \sigma_x=\left(\begin{array}{cc}0 & 1\\ 1 & 0\end{array}\right),$$
$$\sigma_y=\left(\begin{array}{cc}0 & -i\\ i & 0\end{array}\right),\hspace{2em} \sigma_z=\left(\begin{array}{cc}1 & 0\\ 0 & -1\end{array}\right).$$

\item Let $k$ be an integer. Then the $k-$translation operator on
$\mathcal{H}_\infty$ is defined by
$$U_{+k}|n\rangle=|n+k\rangle$$ for all $n\in\mathbb{Z}$. It is easy
to verify that $U_{+k}$ is a unitary operator.
\end{enumerate}
\end{exam}

The basic postulate of quantum mechanics about evolution of systems
may be stated as follows: Suppose that the states of a closed
quantum system at times $t_0$ and $t$ are $|\psi_0\rangle$ and
$|\psi\rangle$, respectively. Then they are related to each other by
a unitary operator $U$ which depends only on the times $t_0$ and
$t$,
$$|\psi\rangle=U|\psi_0\rangle.$$ This postulate can be reformulated
in the language of density operators as follows: The state $\rho$ of
a closed quantum system at time $t$ is related to its state $\rho_0$
at time $t_0$ by a unitary operator $U$ which depends only on the
times $t$ and $t_0$, $$\rho=U\rho_0 U^{\dag}.$$

\subsection{Quantum Measurements}

A quantum measurement on a system with state space $\mathcal{H}$ is
described by a collection $\{M_m\}$ of operators on $\mathcal{H}$
satisfying
$$\sum_{m}M^{\dag}_mM_m=Id_{\mathcal{H}},$$ where $M_m$ are called measurement operators,
and the index $m$ stands for the measurement outcomes that may occur
in the experiment. If the state of a quantum system is
$|\psi\rangle$ immediately before the measurement, then the
probability that result $m$ occurs is
$$p(m)=\langle \psi|M_m^{\dag}M_m|\psi\rangle$$ and the state of the
system after the measurement is
$$|\psi_m\rangle=\frac{M_m|\psi\rangle}{\sqrt{p(m)}}.$$ We can also
formulate the quantum measurement postulate in the language of
density operators. If the state of a quantum system was $\rho$
immediately before measurement $\{M_m\}$ is performed on it, then
the probability that result $m$ occur is
$$p(m)=tr(M_m^{\dag}M_m\rho),$$ and the state of the system after
the measurement is $$\rho_m=\frac{M_m\rho M_m^{\dag}}{p(m)}.$$

A special class of quantum measurements will be frequently used in
the sequel: If a measurement $M$ has only two outcomes, say $0$ and
$1$; that is, $M=\{M_0,M_1\}$, then we often call $M$ a yes-no
measurement, with $0$ corresponding to \textquotedblleft
no\textquotedblright and $1$ to \textquotedblleft
yes\textquotedblright.

\subsection{Tensor Products of Hilbert Spaces}

The state space of a composite quantum system is the tensor product
of the state spaces of its components. In this subsection, we recall
the definition of the tensor product of a family $\{\mathcal{H}_i\}$
of Hilbert spaces. For simplicity of presentation, it will be
assumed that the set of quantum variables is countably infinite, and
the type of each quantum variable is either $\mathbf{Boolean}$ or
$\mathbf{integer}$ (see next section). Thus, we only need to
consider a finite or countably infinite family $\{\mathcal{H}_i\}$
where each $\mathcal{H}_i$ is finite-dimensional or countably
infinite-dimensional. A more general notion of tensor product
introduced by von Neumann~\cite{vN38} should be adopted in order to
generalize the results obtained in this paper to the case of more
quantum variables and other types.

Let $\{|\psi_{ij_i}\rangle\}$ be an orthonormal basis of
$\mathcal{H}_i$ for each $i$. We write $\mathcal{B}$ for the set of
tensor products of basis vectors of all $\mathcal{H}_i$; that is,
$$\mathcal{B}=\{\bigotimes_i|\psi_{ij_i}\rangle\}.$$ Then
$\mathcal{B}$ is a countably infinite set, and it can be written in
the form of a sequence of vectors:
$$\mathcal{B}=\{|\varphi_n\rangle:n=0,1,...\}.$$ The tensor product
of $\{\mathcal{H}_i\}$ is defined to be the Hilbert space spanned by
$\mathcal{B}$, i.e.
$$\bigotimes_i\mathcal{H}_i=\{\sum_n|\varphi_n\rangle:\alpha_n\in\mathbb{C}\ {\rm for\ all}\ n\geq 0\ {\rm and}\
\sum_n|\alpha_n|^{2}<\infty\}.$$ We define the inner product in
$\bigotimes_i\mathcal{H}_i$ as follows:
$$(\sum_n\alpha_n|\psi_n\rangle,\sum_n\alpha_n^{\prime}|\psi_n\rangle)
=\sum_n\alpha_n^{\ast}\alpha_n^{\prime}$$ for any
$\alpha_n,\alpha_n^{\prime}\in\mathbb{C}$, $n\geq 0$. It is easy to
see that $\bigotimes_i\mathcal{H}_i$ is isomorphic to
$\mathcal{H}_\infty$.

The notion of partial trace is very useful for description of a
subsystem of a composite quantum system. Let $\mathcal{H}$ and
$\mathcal{K}$ be two Hilbert spaces and operator
$A\in\mathcal{L}(\mathcal{H}\otimes\mathcal{K})$. Then the partial
trace of $A$ on $\mathcal{H}$ is defined to be
$$tr_\mathcal{K}(A)=\sum_i(I_\mathcal{H}\otimes\langle\psi_i|)\rho(I_\mathcal{H}\otimes
|\psi_i\rangle),$$ which is an operator on $\mathcal{H}$, where
$\{|\psi_i\rangle\}$ is an orthonormal basis of $\mathcal{K}$. It
can be shown that $tr_\mathcal{K}(A)$ does not depend on choice of
$\{|\psi_i\rangle\}$. In particular, if $\mathcal{H}_1$ and
$\mathcal{H}_2$ are the state spaces of quantum systems $q_1$ and
$q_2$, respectively, and the state of their composite system
$q_1q_2$ is described by a density operator
$\rho\in\mathcal{D}^{-}(\mathcal{H}_1\otimes\mathcal{H}_2)$, then
$tr_{\mathcal{H}_2}(\rho)$ is the description for the state of
component system $q_1$.

\section{Syntax of Quantum Programs}

We assume a countably infinite set $Var$ of quantum variables. The
symbols $q, q^{\prime},q^{\prime\prime},$ $q_0,q_1,q_2,...$ will be
used as meta-variables ranging over quantum variables. Recall that
in classical computation, we use a type to denote the domain of a
variable. Thus, in quantum computation, a type should be the state
space of a quantum system denoted by some quantum variable.
Formally, a type $t$ is a name of a Hilbert space $\mathcal{H}_t$.
In this paper, we only consider two basic types: $\mathbf{Boolean},\
\mathbf{integer}$. The results obtained in this paper can be easily
generalized to the case with more types. The Hilbert spaces denoted
by $\mathbf{Boolean}$ and $\mathbf{integer}$ are:
$$\mathcal{H}_{\mathbf{Boolean}}=\mathcal{H}_2,$$
$$\mathcal{H}_{\mathbf{integer}}=\mathcal{H}_\infty.$$
Note that the sets denoted by types $\mathbf{Boolean}$ and
$\mathbf{integer}$ in classical computation are exactly the
computational bases of $\mathcal{H}_{\mathbf{Boolean}}$ and
$\mathcal{H}_{\mathbf{integer}}$, respectively (see
Example~\ref{ex-qub}). Now we assume that each quantum variable $q$
has a type $type(q)$, which is either $\mathbf{Boolean}$ or
$\mathbf{integer}$. The state space $\mathcal{H}_q$ of a quantum
variable $q$ is the Hilbert space denoted by its type; that is,
$$\mathcal{H}_q=\mathcal{H}_{type(q)}.$$

A quantum register is defined to be a finite sequence of distinct
quantum variables. The state space of a quantum register
$\overline{q}=q_1,...,q_n$ is the tensor product of the state spaces
of the quantum variables occurring in $\overline{q}$; that is,
$$\mathcal{H}_{\overline{q}}=\bigotimes_{i=1}^{n}\mathcal{H}_{q_i}.$$

Now we are able to define the syntax of quantum programs. The
quantum programs considered in this paper are quantum extension of
classical \textbf{while}-programs. Formally, they are generated by
the following grammar:
$$S::=\mathbf{skip}\ |\ q:=0\ |\ \overline{q}:=U\overline{q}\ |\ S_1;S_2\ |\ \mathbf{measure}\ M[\overline{q}]:\overline{S}
\ |\ \mathbf{while}\ M[\overline{q}]=1\ \mathbf{do}\ S$$ where
\begin{itemize}\item\ $q$ is a quantum variable and $\overline{q}$ a
quantum register;
\item\ $U$ in the statement \textquotedblleft $\overline{q}:=U\overline{q}$\textquotedblright is a unitary operator on
$\mathcal{H}_{\overline{q}}$. In particular, if
$type(q)=\mathbf{integer}$, then the statement $q:=U_{+k}q$, where
$U_{+k}$ is the $k-$translation operator, will be often abbreviated
to $q:=q+k$;\item\ in the statement \textquotedblleft
$\mathbf{measure}\ M[\overline{q}]:\overline{S}$\textquotedblright,
$M=\{M_m\}$ is a measurement on the state space
$\mathcal{H}_{\overline{q}}$ of $\overline{q}$, and $S=\{S_m\}$ is a
set of quantum programs such that each outcome $m$ of measurement
$M$ corresponds to $S_m$;\item\ $M=\{M_0,M_1\}$ in the statement
\textquotedblleft $\mathbf{while}\ M[\overline{q}]=1\ \mathbf{do}\
S$\textquotedblright is a yes-no measurement on
$\mathcal{H}_{\overline{q}}$.
\end{itemize}

The intuitive meaning of these quantum program constructs will
become clear after introducing their operational semantics in the
next section.

The following technical definition will be needed in the sequel.

\begin{defn}The set $var(S)$ of quantum variables in quantum
program $S$ is recursively defined as follows:\begin{enumerate}
\item If $S=\mathbf{skip}$, then $var(S)=\emptyset$;\item If
$S=q:=0$, then $var(S)=\{q\}$;\item If
$S=\overline{q}:=U\overline{q}$, then $var(S)=\{\overline{q}\}$;
\item If $S=S_1;S_2$, then $var(S)=var(S_1)\cup var(S_2)$; \item If
$S=\mathbf{measure}\ M[\overline{q}]:\overline{S}$, then
$$var(S)=\{\overline{q}\}\cup \bigcup_{m}var(S_m);$$ \item If
$S=\mathbf{while}\ M[\overline{q}]=1\ \mathbf{do}\ S$, then
$var(S)=\{\overline{q}\}\cup var(S).$
\end{enumerate}
\end{defn}

\section{Operational Semantics of Quantum Programs}

We write $\mathcal{H}_{{\rm all}}$ for the tensor product of the
state spaces of all quantum variables, that is,
$$\mathcal{H}_{{\rm all}}=\bigotimes_{{\rm all}\ q}\mathcal{H}_q.$$
For simplicity of presentation, we will use $E$ to denote the empty
program. A quantum configuration is a pair $\langle S,\rho\rangle$,
where $S$ is a quantum program or $E$,
$\rho\in\mathcal{D}^{-}(\mathcal{H}_{{\rm all}})$ is a partial
density operator on $\mathcal{H}_{{\rm all}}$, and it is used to
indicate the (global) state of quantum variables.

Let $\overline{q}=q_1,...,q_n$ be a quantum register. A linear
operator $A$ on $\mathcal{H}_{\overline{q}}$ has a cylinder
extension \begin{equation}\label{cyl}A\otimes
I_{Var-\{\overline{q}\} }\end{equation} on $\mathcal{H}_{{\rm
all}}$, where $I_{Var-\{\overline{q}\}}$ is the identity operator on
the Hilbert space
$$\bigotimes_{q\in Var-\{\overline{q}\}}\mathcal{H}_q.$$ In the
sequel, we will simply for $A$ for its extension~(\ref{cyl}), and it
can be easily recognized from the context, without any risk of
confusion.

The operational semantics of quantum program is defined to be a
transition relation $\rightarrow$ between quantum configurations. By
a transition $$\langle S,\rho\rangle\rightarrow\langle
S^{\prime},\rho^{\prime}\rangle$$ we mean that after executing
quantum program $S$ one step in state $\rho$, the state of quantum
variables becomes $\rho^{\prime}$, and $S^{\prime}$ is the remainder
of $S$ still to be executed. In particular, if $S^{\prime}=E$, then
$S$ terminates in state $\rho^{\prime}$. The transition relation
$\rightarrow$ is given by the transition rules in Fig.1.

\begin{figure}[h]\centering
\begin{equation*}\begin{split}&(Skip)\ \ \ \ \ \ \ \ \
\frac{}{\langle\mathbf{skip},\rho\rangle\rightarrow\langle E,\rho\rangle}\\
&\\ &(Initialization)\ \ \ \ \ \ \ \ \ \frac{}{\langle
q:=0,\rho\rangle\rightarrow\langle E,\rho^{q}_0\rangle}\\ \\& {\rm
where}\\ &\ \ \ \ \ \ \ \ \ \ \ \ \ \ \ \ \ \ \ \ \ \ \ \ \
\rho^{q}_0=|0\rangle_q\langle 0|\rho|0\rangle_q\langle
0|+|0\rangle_q\langle 1|\rho|1\rangle_q\langle 0|\\ & {\rm if}\
type(q)=\mathbf{Boolean},\ {\rm and}
\\ &\ \ \ \ \ \ \ \ \ \ \ \ \ \ \ \ \ \ \ \ \ \ \ \ \ \rho^{q}_0=\sum_{n=-\infty}^{\infty}|0\rangle_q\langle n|\rho|n\rangle_q\langle
0|\\ &{\rm if}\ type(q)=\mathbf{integer}.\\ &\\
&(Unitary\ Transformation)\ \ \ \ \ \ \ \ \
\frac{}{\langle\overline{q}:=U\overline{q},\rho\rangle\rightarrow\langle
E,U\rho U^{\dag}\rangle}\\ &\\ &(Sequential\ Composition)\ \ \ \ \ \
\ \ \ \frac{\langle S_1,\rho\rangle\rightarrow\langle
S_1^{\prime},\rho^{\prime}\rangle} {\langle
S_1;S_2,\rho\rangle\rightarrow\langle
S_1^{\prime};S_2,\rho\rangle}\\ \\&{\rm where\ we\ make\ the\
convention\
that}\ E;S_2=S_2.\\
&\\
&(Measurement)\ \ \ \ \ \ \ \ \ \frac{}{\langle\mathbf{measure}\
M[\overline{q}]:\overline{S},\rho\rangle\rightarrow\langle
S_m,M_m\rho M_m^{\dag}\rangle}\\ &\\ &{\rm for\ each\ outcome}\ m\ {\rm of\ measurement}\ M=\{M_m\}\\
&\\ &(Loop\ 0)\ \ \ \ \ \ \ \ \ \frac{}{\langle\mathbf{while}\
M[\overline{q}]=1\ \mathbf{do}\
S,\rho\rangle\rightarrow\langle E, M_0\rho M_0^{\dag}\rangle}\\ &\\
&(Loop\ 1)\ \ \ \ \ \ \ \ \ \frac{}{\langle\mathbf{while}\
M[\overline{q}]=1\ \mathbf{do}\ S,\rho\rangle\rightarrow\langle
S;\mathbf{while}\ M[\overline{q}]=1\ \mathbf{do}\ S, M_1\rho
M_1^{\dag}\rangle}
\end{split}\end{equation*}
\caption{Transitional Semantics of Quantum Programs}\label{fig 1}
\end{figure}

The meanings of various program constructs are precisely specified
by the transitional rules in Fig.1. The statement \textquotedblleft
$\mathbf{skip}$\textquotedblright does nothing and terminates
immediately. The initialization \textquotedblleft
$q:=0$\textquotedblright sets quantum variable $q$ to the basis
state $|0\rangle$. To see the role of initialization more clearly,
we consider the case of $type(q)=\mathbf{integer}$ as an example.
First, suppose $\rho$ is a pure state; that is,
$\rho=|\psi\rangle\langle\psi|$ for some
$|\psi\rangle\in\mathcal{H}_{{\rm all}}$. We can write
$|\psi\rangle$ in the form: $$|\psi\rangle=\sum_k \alpha_k
|\psi_k\rangle,$$ where $|\psi_k\rangle$ is a product state, say
$$|\psi_k\rangle=\bigotimes_{{\rm all}\
q^{\prime}}|\psi_{kq^{\prime}}\rangle.$$ Then
$$\rho=\sum_{k,l}\alpha_k\alpha_l^{\ast}|\psi_k\rangle\langle\psi_l|.$$ After the initialization the state
becomes:

\begin{equation}\label{init}\begin{split}
\rho^{q}_0&=\sum_{n=-\infty}^{\infty}|0\rangle_q\langle
n|\rho|n\rangle_n\langle 0|\\
&=\sum_{k,l}\alpha_k\alpha_l^{\ast}(\sum_{n=-\infty}^{\infty}|0\rangle_q\langle
n|\psi_k\rangle\langle\psi_l|n\rangle_q\langle 0|)\\
&=\sum_{k,l}\alpha_k\alpha_l^{\ast}(\sum_{n=-\infty}^{\infty}\langle\psi_{lq}|n\rangle\langle
n|\psi_{kq}\rangle)(|0\rangle_q\langle 0|\otimes
\bigotimes_{q^{\prime}\neq
q}|\psi_{kq^{\prime}}\rangle\langle\psi_{lq^{\prime}}|)\\
&=\sum_{k,l}\alpha_k\alpha_l^{\ast}\langle\psi_{lq}|\psi_{kq}\rangle(|0\rangle_q\langle
0|\otimes \bigotimes_{q^{\prime}\neq
q}|\psi_{kq^{\prime}}\rangle\langle\psi_{lq^{\prime}}|)\\
&=|0\rangle_q\langle 0|\otimes
(\sum_{k,l}\alpha_k\alpha_l^{\ast}\langle\psi_{lq}|\psi_{kq}\rangle\bigotimes_{q^{\prime}\neq
q}|\psi_{kq^{\prime}}\rangle\langle\psi_{lq^{\prime}}|).
\end{split}\end{equation}In general, suppose $\rho$ is generated by
an ensemble $\{(p_i,|\psi_i\rangle)\}$ of pure states, that is,
$$\rho=\sum_i p_i|\psi_i\rangle\langle\psi_i|.$$ For each $i$, we
write $\rho_i=|\psi_i\rangle\langle\psi_i|$ and assume that it
becomes $\rho^{q}_{i0}$ after the initialization. By the above
argument, we can write $\rho_{i0}$ in the form:
$$\rho^{q}_{i0}=\sum_k\alpha_{ik}(|0\rangle_q\langle 0|\otimes
|\varphi_{ik}\rangle\langle\varphi_{ik}|),$$ where
$|\varphi_{ik}\rangle\in\mathcal{H}_{var-\{q\}}$ for all $k$. Then
the initialization makes that $\rho$ becomes
\begin{equation}\label{init1}\begin{split}
\rho^{q}_0&=\sum_{n=-\infty}^{\infty}|0\rangle_q\langle
n|\rho|n\rangle_q\langle 0|\\
&=\sum_i p_i(\sum_{n=-\infty}^{\infty}|0\rangle_q\langle
n|\rho_i|n\rangle_q\langle 0|)\\
&=\sum_{i,k}p_i\alpha_{ik}(|0\rangle_q\langle 0|\otimes
|\varphi_{ik}\rangle\langle\varphi_{ik}|).
\end{split}\end{equation}From Eqs.~(\ref{init}) and (\ref{init1}) we
see that the state of $q$ is set to be $|0\rangle$ and the states of
the other quantum variables are unchanged. The statement
\textquotedblleft $\overline{q}:=U\overline{q}$\textquotedblright
simply means that unitary transformation $U$ is performed on quantum
register $\overline{q}$, leaving the states of the quantum variables
not in $\overline{q}$ unchanged. Remark that $U$ in the target
configuration of the rule (Unitary Transformation) stands indeed for
the cylinder extension of $U$ on $\mathcal{H}_{all}$ (see
Eq.~(\ref{cyl})). A similar remark applies to the rules for
measurements and loops. Sequential composition is similar to its
counterpart in classical computation. The program construct
\textquotedblleft $\mathbf{measure}\
M[\overline{q}]:\overline{S}$\textquotedblright is a quantum
generalization of classical conditional statement. Recall that the
first step of the execution of conditional statement
\textquotedblleft $\mathbf{if}\ B\ \mathbf{then}\ S_1\
\mathbf{else}\ S_2\ \mathbf{fi}$\textquotedblright is to check
whether Boolean expression $B$ is satisfied. However, according to a
basic postulate of quantum mechanics, the unique way to acquire
information about a quantum system is to perform a measurement on
it. So, in executing the statement \textquotedblleft
$\mathbf{measure}\ M[\overline{q}]:\overline{S}$\textquotedblright,
quantum measurement $M$ will first be performed on quantum register
$\overline{q}$, and then a subprogram $S_m$ in $\overline{S}$ will
be selected to be executed next according to the outcome of
measurement. The essential difference between a measurement
statement and a classical conditional statement is that the state of
program variables is changed after performing the measurement in the
former, whereas it is not changed after checking the Boolean
expression in the latter. Note that the outcome $m$ is observed with
probability
$$p_m=tr(M_m\rho M_m^{\dag}),$$ and after the measurement the state
becomes $$\rho_m=M_m\rho M_m^{\dag}/p_m.$$ So, a natural
presentation of the Measurement rule is the probabilistic
transition: $$\frac{}{\langle\mathbf{measure}\
M[\overline{q}]:\overline{S},\rho\rangle\stackrel{p_m}{\rightarrow}\langle
S_m,\rho_m\rangle}$$ However, we adopt Selinger's
suggestion~\cite{Se04} of encoding both probability $p_m$ and
density operator $\rho_m$ into partial density operator $$M_m\rho
M_m^{\dag}=p_m\rho_m.$$ This allows us to give the Measurement rule
in terms of ordinary transition. The statement \textquotedblleft
$\mathbf{while}\ M[\overline{q}]=1\ \mathbf{do}\
S$\textquotedblright is a quantum generalization of classical loop
\textquotedblleft$\mathbf{while}\ B\ \mathbf{do}\ S\
\mathbf{od}$\textquotedblright. To acquire information about quantum
register $\overline{q}$, a measurement $M$ is performed on it. The
measurement $M$ is a yes-no measurement with only two possible
outcomes $0,1$. If the outcome $0$ (no) is observed, then the
program terminates, and if the outcome $1$ (yes) occurs, then the
program executes the subprogram $S$ and continues. The only
difference between a quantum loop and a classical loop is that
checking the loop guard $B$ in a classical loop does not change the
state of program variables, but in a quantum loop the measurement
outcomes $0$ and $1$ occur with probabilities:
$$p_0=tr(M_0\rho M_0^{\dag}),\ \ p_1=tr(M_1\rho M_1^{\dag}),$$
respectively, and the state becomes $M_0\rho M_0^{\dag}$ from $\rho$
when the outcome is $0$, and it becomes $M_1\rho M_1^{\dag}$ when
the outcome is $1$. Again, we adopt Selinger's suggestion so that
the (Loop 0) and (Loop 1) rules can be stated as ordinary
transitions instead of probabilistic transitions.

Let $S$ be a quantum program and
$\rho\in\mathcal{D}^{-}(\mathcal{H})$. If $\langle
S^{\prime},\rho^{\prime}\rangle$ can be reached from $\langle
S,\rho\rangle$ in $n$ steps in the transition relation
$\rightarrow$; that is, there are configurations $\langle
S_1,\rho_1\rangle, ...,$ $\langle S_{n-1},\rho_{n-1}\rangle$ such
that $$\langle S,\rho\rangle\rightarrow\langle
S_1,\rho_1\rangle\rightarrow ...\rightarrow\langle
S_{n-1},\rho_{n-1}\rangle\rightarrow \langle
S^{\prime},\rho^{\prime}\rangle,$$ then we write: $$\langle
S,\rho\rangle\rightarrow^{n}\langle
S^{\prime},\rho^{\prime}\rangle.$$ A transition sequence of $S$
starting in $\rho$ is a finite or infinite sequence of
configurations in the following form: $$\langle
S,\rho\rangle\rightarrow\langle S_1,\rho_1\rangle\rightarrow ...
\rightarrow\langle S_n,\rho_n\rangle\rightarrow\langle
S_{n+1},\rho_{n+1}\rangle\rightarrow....
$$ If it cannot be extended, then it is called a computation of $S$
starting in $\rho$. Moveover, if it is finite and its last
configuration is $\langle E,\rho^{\prime}\rangle$, then we say that
it terminates in $\rho^{\prime}$; and if it is infinite, then we say
that it diverges. We say that $S$ can diverge from $\rho$ whenever
it has a diverging computation starting in $\rho$.

Classical \textbf{while}-programs are a typical class of
deterministic programs that have exactly one computation starting in
a given state. As shown in the following example, however, quantum
\textbf{while}-programs no longer possess such a determinism because
probabilism is introduced by the measurements in the statements
\textquotedblleft$\mathbf{measure}\
M[\overline{q}]:\overline{S}$\textquotedblright and
\textquotedblleft$\mathbf{while}\ M[\overline{q}]=1\ \mathbf{do}\
S$\textquotedblright. After encoding probabilities into partial
density operators, probabilism manifests as nondeterminism in
transition rules (Measurement), (Loop 0) and (Loop 2).

\begin{exam}Suppose that $type(q_1)=\mathbf{Boolean}$ and
$type(q_2)=\mathbf{integer}$. Consider the program:
$$S=q_1:=0;q_2:=0;q_1:=Hq_1;q_2:=q_2+2;\mathbf{measure}\
M[q_1]:\overline{S}$$ where \begin{itemize}\item\ $M$ is the
measurement according to the computational basis
$\{|0\rangle,|1\rangle\}$ of $\mathcal{H}_2$; that is, $M=\{M_0,
M_1\}$, $M_0=|0\rangle\langle 0|$ and $M_1=|1\rangle\langle 1|$;
\item\ $\overline{S}=S_1,S_2$, and \begin{itemize}\item\
$S_1=\mathbf{skip}$; \item\ $S_2=\mathbf{while}\ N[q_2]=1\
\mathbf{do}\ q_1:=\sigma_zq_1$, where $N=\{N_0,N_1\}$,
$$N_0=\sum_{n=-\infty}^{0}|n\rangle\langle n|\ {\rm and}\
N_1=\sum_{n=1}^{\infty}|n\rangle\langle n|.$$
\end{itemize}\end{itemize}
Let $$\rho_0=\bigotimes_{q\neq q_1,q_2}|0\rangle_q\langle 0|$$ and
$$\rho=|1\rangle_{q_1}\langle 1|\otimes |-1\rangle_{q_2}\langle
-1|\otimes\rho_0.$$ Then the computations of $S$ starting in $\rho$
are:\begin{equation*}\begin{split} \langle
S,\rho\rangle&\rightarrow\langle
q_2:=0;q_1:=Hq_1;q_2:=q_2+2;\mathbf{measure},\rho_1\rangle\\
&\rightarrow\langle q_1:=Hq_1;q_2:=q_2+2;\mathbf{measure},\rho_2\rangle\\
&\rightarrow\langle q_2:=q_2+2;\mathbf{measure},\rho_3\rangle\\
&\rightarrow\langle \mathbf{measure},\rho_4\rangle\\
&\rightarrow\begin{cases}\langle S_1,\rho_5\rangle\rightarrow\langle
E,\rho_5\rangle,\\ \langle S_2,\rho_6\rangle,
\end{cases}
\end{split}
\end{equation*}
\begin{equation*}
\begin{split}\langle S_2,\rho_6\rangle&\rightarrow\langle q_1:=\sigma_zq_1;S_2,\rho_6\rangle\\
&\rightarrow \langle S_2, -\rho_6\rangle\\ &\rightarrow ... \\
&\rightarrow^{2n-1}\langle q_1:=\sigma_zq_1;S_2,(-1)^{n-1}\rho_6\rangle\\
&\rightarrow \langle S_2, (-1)^{n}\rho_6\rangle\\ &\rightarrow ...
\end{split}\end{equation*}
where $\mathbf{measure}$ stands for the statement
\textquotedblleft$\mathbf{measure}\
M[q_1]:\overline{S}$\textquotedblright, and
\begin{equation*}\begin{split} \rho_1&=|0\rangle_{q_1}\langle
0|\otimes |-1\rangle_{q_2}\langle -1|\otimes\rho_0,\\
\rho_2&=|0\rangle_{q_1}\langle
0|\otimes |0\rangle_{q_2}\langle 0|\otimes\rho_0,\\
\rho_3&=|+\rangle_{q_1}\langle
+|\otimes |0\rangle_{q_2}\langle 0|\otimes\rho_0,\\
\rho_4&=|+\rangle_{q_1}\langle +|\otimes |2\rangle_{q_2}\langle
2|\otimes\rho_0,
\\
\rho_5&=\frac{1}{2}|0\rangle_{q_1}\langle 0|\otimes
|2\rangle_{q_2}\langle
2|\otimes\rho_0,\\
\rho_6&=\frac{1}{2}|1\rangle_{q_1}\langle 1|\otimes
|2\rangle_{q_2}\langle 2|\otimes\rho_0.
\end{split}\end{equation*}
So, $S$ can diverge from $\rho$. Note that $S_2$ has also the
transition $$\langle S_2,(-1)^{n}\rho_6\rangle\rightarrow\langle
E,0_{\mathcal{H}_{all}}\rangle,$$ but we always discard the
transitions in which the partial density operator of the target
configuration is zero operator.
\end{exam}

\section{Denotational Semantics of Quantum Programs}

The denotational semantics of a quantum program is defined to be a
semantic function which maps partial density operators to
themselves. More precisely, for any quantum program $S$, the
semantic function of $S$ sums the computated results of all
terminating computations of $S$.

We write $\rightarrow^{\ast}$ for the reflexive and transitive
closures of $\rightarrow$; that is, $\langle
S,\rho\rangle\rightarrow^{\ast}\langle
S^{\prime},\rho^{\prime}\rangle$ if and only if $\langle
S,\rho\rangle\rightarrow^{n}\langle S^{\prime},\rho^{\prime}\rangle$
for some $n\geq 0$.

\begin{defn}\label{semdef}Let $S$ be a quantum program. Then its semantic function $$[|S|]:
\mathcal{D}^{-}(\mathcal{H}_{{\rm
all}})\rightarrow\mathcal{D}^{-}(\mathcal{H}_{{\rm all}})$$ is
defined by
\begin{equation}\label{denot}[|S|](\rho)=\sum\{|\rho^{\prime}:\langle
S,\rho\rangle\rightarrow^{\ast}\langle
E,\rho^{\prime}\rangle|\}\end{equation} for all
$\rho\in\mathcal{H}_{{\rm all}}$.
\end{defn}

It should be pointed out that $\{|\cdot|\}$ in Eq.~(\ref{denot})
stands for multi-set. The reason for using multi-sets is that the
same density operator may be obtained through different
computational paths as we can see from the measurement and loop
rules in the operational semantics. The following simple example
illustrates the case more explicitly.

\begin{exam}Assume that $type(q)=\mathbf{Boolean}$. Consider the
program: $$S=q:=0;q:=Hq;\mathbf{measure}\ M[q]:\overline{S},$$ where
\begin{itemize}\item\ $M$ is the measurement according to the
computational basis $\{|0\rangle,|1\rangle\}$ of $\mathcal{H}_2$;
\item\ $\overline{S}=S_0,S_1$, $S_0=q:=Iq$ and $S_1=q:=\sigma_x q$.
\end{itemize}Let $\rho=|0\rangle_{{\rm all}}\langle 0|$, where $$|0\rangle_{{\rm all}}=\bigotimes_{{\rm all}\
q}|0\rangle_q.$$ Then the computation of $S$ starting in $\rho$ is
given as follows:\begin{equation*}\begin{split}\langle
S,\rho\rangle&\rightarrow \langle
q:=Hq;\mathbf{measure},\rho\rangle\\
&\rightarrow\langle\mathbf{measure},|+\rangle_q\langle
+|\otimes\bigotimes_{q^{\prime}\neq q}|0\rangle_{q^{\prime}}\langle
0|\rangle\\ &\rightarrow\begin{cases} \langle
S_0,\frac{1}{2}|0\rangle_q\langle
0|\otimes\bigotimes_{q^{\prime}\neq q}|0\rangle_{q^{\prime}}\langle
0|\rangle\rightarrow\langle E,\frac{1}{2}\rho\rangle,\\
\langle S_1,\frac{1}{2}|1\rangle_q\langle
1|\otimes\bigotimes_{q^{\prime}\neq q}|0\rangle_{q^{\prime}}\langle
0|\rangle\rightarrow\langle E,\frac{1}{2}\rho\rangle,
\end{cases}
\end{split}\end{equation*}where $\mathbf{measure}$ is an
abbreviation of \textquotedblleft $\mathbf{measure}\
M[q]:\overline{S}$\textquotedblright. So, we have:
$$[|S|](\rho)=\frac{1}{2}\rho+\frac{1}{2}\rho=\rho.$$
\end{exam}

Now we are going to establish some basic properties of semantic
functions. First, we prove its linearity.

\begin{lem}\label{linear}
Let $\rho_1,\rho_2\in\mathcal{D}^{-}(D)$ and
$\lambda_1,\lambda_2\geq 0$. If
$\lambda_1\rho_1+\lambda_2\rho_2\in\mathcal{D}^{-}(D)$, then for any
quantum program $S$, we have:
$$[|S|](\lambda_1\rho_1+\lambda_2\rho_2)=\lambda_1 [|S|](\rho_1)+\lambda_2 [|S|](\rho_2).$$
\end{lem}

\textit{Proof.} We can easily prove the following fact by induction
on the structure of $S$:\begin{itemize}\item \ Claim: If $\langle
S,\rho_1\rangle\rightarrow \langle
S^{\prime},\rho_1^{\prime}\rangle$ and $\langle
S,\rho_2\rangle\rightarrow \langle
S^{\prime},\rho_2^{\prime}\rangle$, then $$\langle
S,\lambda_1\rho_1+\lambda_2\rho_2\rangle\rightarrow \langle
S^{\prime},\lambda_1\rho_1^{\prime}+\lambda_2\rho_2^{\prime}\rangle.$$
\end{itemize}Then the conclusion immediately follows. $\Box$

Next we give a representation of semantic function $[|S|]$ according
to the structure of program $S$. To do this for quantum loops, we
need some auxiliary notations. Let $\Omega$ be a quantum program
such that $[|\Omega|]=0_{\mathcal{H}_{{\rm all}}}$ for all
$\rho\in\mathcal{D}(\mathcal{H})$; for example,
$$\Omega=\mathbf{while}\ M_{{\rm trivial}}[q]=1\ \mathbf{do\
skip},$$ where $q$ is a quantum variable, and $$M_{{\rm
trivial}}=\{M_0=0_{\mathcal{H}_{q}},M_1=I_{\mathcal{H}_{q}}\}$$ is a
trivial measurement on $\mathcal{H}_{q}.$ We
set:\begin{equation*}\begin{split} (\mathbf{while}\
M[\overline{q}]=1\ \mathbf{do}\ S)^{0}&=\Omega,\\
(\mathbf{while}\ M[\overline{q}]=1\ \mathbf{do}\
S)^{n+1}&=\mathbf{measure}\ M[\overline{q}]:\overline{S},
\end{split}\end{equation*} where $\overline{S}=S_0,S_1$,
and\begin{equation*}\begin{split}S_0&=\mathbf{skip},\\
S_1&=S;(\mathbf{while}\ M[\overline{q}]=1\ \mathbf{do}\ S)^{n}
\end{split}\end{equation*} for all $n\geq 0$.

\begin{prop}\label{sempro}\begin{enumerate}\item
$[|\mathbf{skip}|](\rho)=\rho$. \item If $type(q)=\mathbf{Boolean}$,
then $$[|q:=0|](\rho)=|0\rangle_q\langle 0|\rho|0\rangle_q\langle
0|+|0\rangle_q\langle 1|\rho|1\rangle_q\langle 0|,$$ and if
$type(q)=\mathbf{integer}$, then
$$[|q:=0|](\rho)\sum_{n=-\infty}^{\infty}|0\rangle_q\langle
n|\rho|n\rangle_q\langle 0|.$$ \item
$[|\overline{q}:=U\overline{q}|](\rho)=U\rho U^{\dag}$.
\item $[|S_1;S_2|](\rho)=[|S_2|]([|S_1|](\rho))$.
\item $[|\mathbf{measure}\
M[\overline{q}]:\overline{S}|](\rho)=\sum_m[|S_m|](M_m\rho
M_m^{\dag})$.
\item $[|\mathbf{while}\ M[\overline{q}]=1\ \mathbf{do}\ S|](\rho)=\bigvee_{n=0}^{\infty}
[|(\mathbf{while}\ M[\overline{q}]=1\ \mathbf{do}\ S)^{n}|](\rho).$
\end{enumerate}\end{prop}

\textit{Proof.} (1), (2) and (3) are obvious.

(4) By Lemma~\ref{linear} and the transitional rule for sequential
composition we obtain:\begin{equation*}\begin{split}
[|S_2|]([|S_1|](\rho))&=[|S_2|](\sum\{|\rho_1:\langle
S_1,\rho\rangle\rightarrow^{\ast}\langle E,\rho_1\rangle|\})\\
&=\sum\{|[|S_2|](\rho_1):\langle
S_1,\rho\rangle\rightarrow^{\ast}\langle E,\rho_1\rangle|\}\\
&=\sum\{|\sum\{|\rho^{\prime}:\langle
S_2,\rho_1\rangle\rightarrow^{\ast}\langle
E,\rho^{\prime}\rangle|\}:\langle
S_1,\rho\rangle\rightarrow^{\ast}\langle E,\rho_1\rangle|\}\\
&=\sum\{|\rho^{\prime}:\langle
S_1,\rho\rangle\rightarrow^{\ast}\langle E,\rho_1\rangle\ {\rm and}\
\langle S_2,\rho_1\rangle\rightarrow^{\ast}\langle
E,\rho^{\prime}\rangle|\}\\ &=\sum\{|\rho^{\prime}:\langle
S_1;S_2,\rho\rangle\rightarrow^{\ast}\langle
E,\rho^{\prime}\rangle|\}\\ &=[|S_1;S_2|](\rho).
\end{split}\end{equation*}

(5) follows immediately from the transitional rule for measurement.

(6) We introduce two auxiliary operators:
$$\mathcal{E}_i:\mathcal{D}^{-}(\mathcal{H}_{{\rm
all}})\rightarrow\mathcal{D}^{-}(\mathcal{H}_{{\rm all}}),$$
$$\mathcal{E}_i(\rho)=M_i\rho M_i^{\dag}$$ for all
$\rho\in\mathcal{D}^{-}(\mathcal{H})$ and $i=0,1$. For simplicity,
we write $\mathbf{while}$ for \textquotedblleft $\mathbf{while}\
M[\overline{q}]=1\ \mathbf{do}\ S$\textquotedblright. First, we
prove:
$$[|(\mathbf{while})^{n}|](\rho)=\sum_{k=0}^{n-1}[\mathcal{E}_0\circ ([|S|]\circ
\mathcal{E}_1)^{k}](\rho)$$ for all $n\geq 1$ by induction on $n$.
The case of $n=1$ is obvious. Then by (1), (4) and (5) and the
induction hypothesis on $n-1$ we obtain:
\begin{equation}\label{sem-fun}\begin{split}
[|(\mathbf{while})^{n}|](\rho)&=[|\mathbf{skip}|](\mathcal{E}_0(\rho))+[|S;(\mathbf{while})^{n}|](\mathcal{E}_1(\rho))\\
&=\mathcal{E}_0(\rho)+[|(\mathbf{while})^{n-1}|](([|S|]\circ\mathcal{E}_1)(\rho))\\
&=\mathcal{E}_0(\rho)+\sum_{k=0}^{n-2}[\mathcal{E}_0\circ([|S|]\circ\mathcal{E}_1)^{k}](([|S|]\circ\mathcal{E}_0)(\rho))\\
&=\sum_{k=0}^{n-1}[\mathcal{E}_0\circ([|S|]\circ\mathcal{E}_1)^{k}](\rho).
\end{split}\end{equation}

Second, we have:
\begin{equation*}\begin{split}
[|\mathbf{while}|](\rho)&=\sum\{|\rho^{\prime}:\langle\mathbf{while},\rho\rangle\rightarrow^{\ast}\langle
E,\rho^{\prime}\rangle|\}\\
&=\sum_{n=1}^{\infty}\sum\{|\rho^{\prime}:\langle\mathbf{while},\rho\rangle\rightarrow^{n}\langle
E,\rho^{\prime}\rangle|\}.
\end{split}\end{equation*}
So, it suffices to show that
$$\sum\{|\rho^{\prime}:\langle\mathbf{while},\rho\rangle\rightarrow^{n}\langle
E,\rho^{\prime}\rangle|\}=[\mathcal{E}_0\circ ([|S|]\circ
\mathcal{E}_1)^{n-1}](\rho)$$ for all $n\geq 1$. This can be easily
done by induction on $n$. $\Box$

If we only consider quantum variables of type \textbf{Boolean}, then
the above proposition coincides with Fig.1 in~\cite{FDJY07}.
However, it is worth noting that in~\cite{FDJY07} the denotational
semantics of quantum programs was directly defined and it lacks a
basis of operational semantics. Similar to Lemma 3.2
in~\cite{FDJY07}, we may prove that the semantic function of a
quantum program is a super-operator. Thus, the denotational
semantics given in this section is consistent with Selinger's idea
of modeling quantum programs as super-operators~\cite{Se04}.

A recursive characterization of the semantic function of a quantum
loop can be derived from the above proposition.

\begin{cor}If we write $\mathbf{while}$ for quantum loop \textquotedblleft$\mathbf{while}\
M[\overline{q}]=1\ \mathbf{do}\ S$\textquotedblright, then for any
$\rho\in\mathcal{D}^{-}(\mathcal{H}_{all})$, it holds that
$$[|\mathbf{while}|](\rho)=M_0\rho M_0^{\dag}+[|\mathbf{while}|]([|S|](M_1\rho
M_1^{\dag})).$$\end{cor}

\textit{Proof.}Immediate from Proposition~\ref{sempro}(6) and
Eq.~(\ref{sem-fun}). $\Box$

The following proposition shows that a semantic function does not
increase the trace of density operator of quantum variables.

\begin{prop}\label{sempro1} For any quantum program $S$, it holds that
$$tr([|S|](\rho))\leq tr(\rho)$$
for all $\rho\in\mathcal{D}^{-}(\mathcal{H}_{{\rm all}})$.
\end{prop}

\textit{Proof.} We proceed by induction on the structure of $S$.

Case 1. $S=\mathbf{skip}$. Obvious.

Case 2. $S=q:=0$. If $type(q)=\mathbf{integer}$, then
\begin{equation*}\begin{split}tr([|S|](\rho))&=\sum_{n=-\infty}^{\infty}tr(|0\rangle_q\langle
n|\rho|n\rangle_n\langle 0|)\\
&=\sum_{n=-\infty}^{\infty}tr(_q\langle 0|0\rangle_q\langle
n|\rho|n\rangle_q)\\
&=tr[(\sum_{n=-\infty}^{\infty}|n\rangle_q\langle n|)\rho]=tr(\rho).
\end{split}\end{equation*}

It can be proved in a similar way when $type(q)=\mathbf{Boolean}$.

Case 3. $S=\overline{q}:=U\overline{q}.$ Then
$$tr([|S|](\rho)=tr(U\rho
U^{\dag})=tr(U^{\dag}U\rho)=tr(\rho).$$

Case 4. $S=S_1;S_2$. It follows from the induction hypothesis on
$S_1$ and $S_2$ that
\begin{equation*}\begin{split}tr([|S|](\rho))&=tr([|S_2|]([|S_1|](\rho)))\\
&\leq tr([|S_1|](\rho))\\ &\leq tr(\rho).\end{split}\end{equation*}

Case 5. $S=\mathbf{measure}\ M[\overline{q}]:\overline{S}$. Then by
induction hypothesis we
obtain:\begin{equation*}\begin{split}tr([|S|](\rho))&=\sum_m
tr([|S_m|](M_m\rho M_m^{\dag}))\\
&\leq \sum_m tr(M_m\rho M_m^{\dag})\\ &=tr[(\sum_m
M_m^{\dag}M_m)\rho]\\ &=tr(\rho).
\end{split}\end{equation*}

Case 6. $S=\mathbf{while}\ M[\overline{q}]=1\ \mathbf{do}\
S^{\prime}$. We write $(\mathbf{while})^{n}$ for statement
\textquotedblleft$(\mathbf{while}\ M[\overline{q}]=1\ \mathbf{do}\
S^{\prime})^{n}$\textquotedblright. With
Proposition~\ref{sempro}(6), it suffices to show that
$$tr([|(\mathbf{while})^{n}|](\rho))\leq tr(\rho)$$ for all $n\geq
0$. This can be carried out by induction on $n$. The case of $n=0$
is obvious. By the induction hypothesis on $n$ and $S^{\prime}$, we
have:\begin{equation*}\begin{split}
tr([|(\mathbf{while})^{n+1}|](\rho))&=tr(M_0\rho
M_0^{\dag})+tr([|(\mathbf{while})^{n}|]_{{\rm
par}}([|S^{\prime}|](M_1\rho M_1^{\dag})))\\
&\leq tr(M_0\rho M_0^{\dag})+tr([|S^{\prime}|](M_1\rho
M_1^{\dag}))\\ &\leq
tr(M_0\rho M_0^{\dag}) + tr(M_1\rho M_1^{\dag})\\
&=tr[(M_0^{\dag}M_0+M_1^{\dag}M_1)\rho]\\ &=tr(\rho).\ \Box
\end{split}\end{equation*}

From the proof of the above proposition, it is easy to see that the
unique possibility that $tr([|S|](\rho))<tr(\rho)$ comes from the
quantum loops occurring in $S$. Thus, $tr(\rho)-tr([|S|](\rho))$ is
the probability that program $S$ diverges from input state $\rho$.
This can be further illustrated by the following example.

\begin{exam}Let $type(q)=\mathbf{integer}$, and let $$M_0=\sum_{n=1}^{\infty}\sqrt{\frac{n-1}{2n}}(|n\rangle\langle n|
+|-n\rangle\langle -n|),$$
$$M_1=\sum_{n=1}^{\infty}\sqrt{\frac{n+1}{2n}}(|n\rangle\langle n|+|-n\rangle\langle -n|)+|0\rangle\langle
0|.$$ Then $M=\{M_0,M_1\}$ is a yes-no measurement on
$\mathcal{H}_q$. Consider the program: $$\mathbf{while}\ M[q]=1\
\mathbf{do}\ q:=q+1.$$ For simplicity, we write $\mathbf{while}$ for
this program. Let $$\rho_0=\bigotimes_{q^{\prime}\neq
q}|0\rangle_{q^{\prime}}\langle 0|$$ and $\rho=|0\rangle_q\langle
0|\otimes\rho_0.$ Then
$$[|(\mathbf{while})^{n}|](\rho)=\begin{cases}0_{\mathcal{H}_{{\rm
all}}}\ &{\rm if}\ n=0,1,2,\\
\frac{1}{2}(\sum_{k=2}^{n-1}\frac{k-1}{k!}|k\rangle_q\langle
k|)\otimes\rho_0\ &{\rm if}\ n\geq 3,
\end{cases}$$
$$[|\mathbf{while}|](\rho)=\frac{1}{2}(\sum_{n=2}^{\infty}\frac{n-1}{n!}|n\rangle_q\langle
n|)\otimes\rho_0,$$ and
$$tr([|\mathbf{while}|](\rho))=\frac{1}{2}\sum_{n=2}^{\infty}\frac{n-1}{n!}=\frac{1}{2}.$$This means
that program $\mathbf{while}$ terminates on input $\rho$ with
probability $\frac{1}{2}$, and it diverges from input $\rho$ with
probability $\frac{1}{2}$.
\end{exam}

To conclude this section, we observe how quantum programs change the
states of quantum variables and how they access quantum variables.

Let $X\subseteq Var$ be a set of quantum variables. For any
$A\in\mathcal{L}(\mathcal{H}_{all})$, we write $tr_X(A)$ for
$$tr_{\bigotimes_{q\in X}\mathcal{H}_q}(A).$$

\begin{prop}\label{change}\begin{enumerate}\item
$tr_{var(S)}([|S|](\rho))=tr_{var(S)}(\rho).$
\item If $tr_{Var-var(S)}(\rho_1)=tr_{Var-var(S)}(\rho_2)$, then
$$tr_{Var-var(S)}([|S|](\rho_1))=tr_{Var-var(S)}([|S|](\rho_2)).$$
\end{enumerate}\end{prop}

We put the long and dumb proof of the above proposition into the
appendix.

Recall that $tr_X(\rho)$ describes the state of the quantum
variables not in $X$ when the global state of all quantum variables
is $\rho$. So, Proposition~\ref{change}(1) indicates that the state
of the quantum variables not in $var(S)$ after implementing quantum
program $S$ is the same as that before implementing $S$. This means
that program $S$ can only change the state of quantum variables in
$var(S)$. On the other hand, Proposition~\ref{change}(2) shows that
if two input states $\rho_1$ and $\rho_2$ coincide on the quantum
variables in $var(S)$, then the computed outcomes of $S$, starting
in $\rho_1$ and $\rho_2$, respectively, will also coincide on these
quantum variables. In other words, program $S$ can access at most
the quantum variables in $var(S)$.

\section{Correctness Formulas}

Correctness of a quantum program will be expressed by a quantum
extension of Hoare triple in which a quantum predicate describes the
input state and a quantum predicate describes the output states of
the program. We adopt D'Hondt and Panangaden's definition of quantum
predicates~\cite{DP06}. For any $X\subseteq Var$, a quantum
predicate on $\mathcal{H}_{X}$ is defined to be a Hermitian operator
$P$ on $\mathcal{H}_{X}$ such that $$0_{\mathcal{H}_{X}}\sqsubseteq
P\sqsubseteq I_{\mathcal{H}_{X}}.$$ We write
$\mathcal{P}(\mathcal{H}_X)$ for the set of quantum predicates on
$\mathcal{H}_X$. Intuitively, for any
$\rho\in\mathcal{D}^{-}(\mathcal{H}_{X})$, $tr(P\rho)$ stands for
the probability that predicate $P$ is satisfied in state $\rho$.

A correctness formula is a statement of the form:
$$\{P\}S\{Q\}$$ where $S$ is a quantum program, and both $P$ and $Q$
are quantum predicates on $\mathcal{H}_{all}$. The quantum predicate
$P$ is called the precondition of the correctness formula and $Q$
the postcondition. A correctness formula can be interpreted in two
different ways:

\begin{defn}\begin{enumerate}\item The correctness formula $\{P\}S\{Q\}$ is true in
the sense of total correctness, written
$$\models_{{\rm tot}}\{P\}S\{Q\},$$ if we have: $$tr(P\rho)\leq tr(Q[|S|](\rho))$$ for all
$\rho\in\mathcal{D}^{-}(\mathcal{H})$.

\item The correctness formula $\{P\}S\{Q\}$ is true in
the sense of partial correctness, written
$$\models_{{\rm par}}\{P\}S\{Q\},$$ if we have: $$tr(P\rho)\leq tr(Q[|S|](\rho))+
[tr(\rho)-tr([|S|](\rho))]
$$ for all $\rho\in\mathcal{D}^{-}(\mathcal{H})$.
\end{enumerate}
\end{defn}

The intuitive meaning of the defining inequality of total
correctness is: the probability that input $\rho$ satisfies quantum
predicate $P$ is not greater than the probability that quantum
program $S$ terminates on $\rho$ and its output $[|S|](\rho)$
satisfies quantum predicate $Q$. Recall that
$tr(\rho)-tr([|S|](\rho))$ is the probability that quantum program
$S$ diverges from input $\rho$. Thus, the definition inequality of
partial correctness means: if input $\rho$ satisfies predicate $P$,
then either program $S$ terminates on it and its output
$[|S|](\rho)$ satisfies $Q$, or $S$ diverges from it. The difference
between total correctness and partial correctness is illustrated
well by the following simple example.

\begin{exam}Assume that $type(q)=\mathbf{Boolean}$. Consider the
program: $$S=\mathbf{while}\ M[q]=1\ \mathbf{do}\ q:=\sigma_z q$$
where $M_0=|0\rangle\langle 0|$ and $M_1=|1\rangle\langle 1|$. Let
$P=|\psi\rangle_q\langle\psi|\otimes P^{\prime}$, where
$|\psi\rangle=\alpha |0\rangle+\beta |1\rangle\in\mathcal{H}_2$, and
$P^{\prime}\in\mathcal{P}(\mathcal{H}_{Var-\{q\}})$. Then
\begin{equation*}\models_{tot}\{P\}S\{|0\rangle_q\langle 0|\otimes
P^{\prime}\}\end{equation*} does not hold if $\beta\neq 0$ and
$P\neq 0_{\mathcal{H}_{Var-\{q\}}}$. In fact, put
$$\rho=|\psi\rangle_q\langle\psi|\otimes
I_{\mathcal{H}_{Var-\{q\}}}.$$ Then
$$[|S|](\rho)=|\alpha|^{2}|0\rangle_q\langle 0|\otimes
I_{\mathcal{H}_{Var-\{q\}}}$$ and
$$tr(P\rho)=tr(P^{\prime})>|\alpha|^{2}tr(P^{\prime})=tr((|0\rangle_q\langle
0|\otimes P^{\prime})[|S|](\rho)).$$ On the other hand, we always
have: \begin{equation*}\models_{par}\{P\}S\{|0\rangle_q\langle
0|\otimes P^{\prime}\}.\end{equation*}To show this, we first
consider a special class of partial density operators on
$\mathcal{H}_{Var-\{q\}}$: $\rho=|\varphi\rangle_q\langle\varphi
|\otimes\rho^{\prime}$, where
$|\varphi\rangle=a|0\rangle+b|1\rangle\in\mathcal{H}_2$, and
$\rho^{\prime}\in\mathcal{D}^{-}(\mathcal{H}_{Var-\{q\}})$. A
routine calculation yields:
$$[|S|](\rho)=|a|^{2}|0\rangle_q\langle 0|\otimes\rho^{\prime}$$
and
\begin{equation*}\begin{split}
tr(P\rho)&=|\langle\psi|\varphi\rangle|^{2}tr(P^{\prime}\rho^{\prime})\\
&\leq
|a|^{2}tr(P^{\prime}\rho^{\prime})+[tr(\rho^{\prime})-|a|^{2}tr(\rho^{\prime})]\\
&=tr((|0\rangle_q\langle 0|\otimes
P^{\prime})[|S|](\rho))+[tr(\rho)-tr([|S|](\rho))].
\end{split}\end{equation*} Then it follows from linearity of $[|S|]$ (Lemma~\ref{linear})
that $$tr(P\rho)\leq tr((|0\rangle_q\langle 0|\otimes
P^{\prime})[|S|](\rho))+[tr(\rho)-tr([|S|](\rho))]$$ for all
$\rho\in\mathcal{D}^{-}(\mathcal{H}_{all})$.
\end{exam}

The following proposition presents some basic properties of
correctness formulas.

\begin{prop}\begin{enumerate}
\item If $\models_{tot}\{P\}S\{Q\}$, then $\models_{par}\{P\}S\{Q\}$.
\item For any quantum program $S$, and for any
$P,Q\in\mathcal{P}(\mathcal{H}_{{\rm all}})$, we have:
$$\models_{{\rm tot}}\{0_{\mathcal{H}_{{\rm all}}}\}S\{Q\},\ \ \models_{{\rm par}}\{P\}S\{I_{\mathcal{H}_{{\rm all}}}\}.$$
\item Linearity: For any $P_1,P_2,Q_1,Q_2\in\mathcal{P}(\mathcal{H}_{{\rm all}})$ and $\lambda_1,\lambda_2\geq 0$ with $\lambda_1 P_1+\lambda_2 P_2,
\lambda_1 Q_1+\lambda_2 Q_2\in\mathcal{P}(\mathcal{H}_{{\rm all}})$,
if $$\models_{tot}\{P_i\}S\{Q_i\}\ (i=1,2),$$ then
$$\models_{tot}\{\lambda_1 P_1+\lambda_2 P_2\}S\{\lambda_1
Q_1+\lambda_2 Q_2\}.$$

The same conclusion holds for partial correctness if
$\lambda_1+\lambda_2=1$.
\end{enumerate}
\end{prop}

\textit{Proof.} Immediate from definition. $\Box$

\section{Weakest Preconditions and Weakest Liberal Preconditions}

As in the case of classical Hoare logic, the notions of weakest
precondition and weakest liberal precondition of quantum program
will play a key role in establishing the (relative) completeness of
Hoare logic for quantum programs. They may be defined in a familiar
way:

\begin{defn}\label{wlp-def}Let $S$ be a quantum program and
$P\in\mathcal{P}(\mathcal{H}_{{\rm all}})$ be a quantum predicate on
$\mathcal{H}_{{\rm all}}$. \begin{enumerate}\item The weakest
precondition of $S$ with respect to $P$ is defined to be the quantum
predicate $wp.S.P\in\mathcal{P}(\mathcal{H}_{{\rm all}})$ satisfying
the following conditions:\begin{enumerate}\item $\models_{{\rm
tot}}\{wp.S.P\}S\{P\};$ \item if quantum predicate
$Q\in\mathcal{P}(\mathcal{H}_{all})$ satisfies $\models_{{\rm
tot}}\{Q\}S\{P\}$ then $Q\sqsubseteq wp.S.P$.
\end{enumerate}
\item The weakest liberal
precondition of $S$ with respect to $P$ is defined to be the quantum
predicate $wlp.S.P\in\mathcal{P}(\mathcal{H}_{{\rm all}})$
satisfying the following conditions:\begin{enumerate}\item
$\models_{{\rm par}}\{wlp.S.P\}S\{P\};$ \item if quantum predicate
$Q\in\mathcal{P}(\mathcal{H}_{all})$ satisfies $\models_{{\rm
par}}\{Q\}S\{P\}$ then $Q\sqsubseteq wlp.S.P$.
\end{enumerate}\end{enumerate}
\end{defn}

The next two propositions give explicit representations of weakest
preconditions and weakest liberal preconditions, respectively. They
will be used in the proof of completeness of quantum Hoare logic.

\begin{prop}\label{wp-re}\begin{enumerate}\item \begin{enumerate}\item $wp.\mathbf{skip}.P=P.$
\item If $type(q)=\mathbf{Boolean}$, then $$wp.q:=0.P=|0\rangle_q\langle
0|P|0\rangle_q\langle 0|+|1\rangle_q\langle 0|P|0\rangle_q\langle
1|,$$ and if $type(q)=\mathbf{integer}$, then
$$wp.q:=0.P=\sum_{n=-\infty}^{\infty} |n\rangle_q\langle 0|P|0\rangle_q\langle
n|.$$ \item $wp.\overline{q}:=U\overline{q}.P=U^{\dag}PU$.
\item $wp.S_1;S_2.P=wp.S_1.(wp.S_2.P)$. \item $wp.\mathbf{measure}\
M[\overline{q}]:\overline{S}.P=\sum_m M_m^{\dag}(wp.S_m.P)M_m.$
\item $wp.\mathbf{while}\ M[\overline{q}]=1\ \mathbf{do}\
S.P=\bigvee_{n=0}^{\infty}P_n,$ where $$\begin{cases}&P_0=0_{\mathcal{H}_{all}},\\
& P_{n+1}=M_0^{\dag}PM_0+M_1^{\dag}(wp.S.P_n)M_1\ {\rm for\ all}\
n\geq 0.
\end{cases}$$
\end{enumerate}\item For any quantum program $S$, for any quantum
predicate $P\in\mathcal{P}(\mathcal{H}_{{\rm all}}),$ and for any
partial density operator $\rho\in\mathcal{D}^{-}(\mathcal{H}_{{\rm
all}})$, we have: $$tr((wp.S.P)\rho)=tr(P[|S|]_{{\rm
par}}(\rho)).$$\end{enumerate}
\end{prop}

\textit{Proof.} The trick is to simultaneously prove (1) and (2) by
induction on the structure. This is indeed why we put these two
conclusions that seem irrelevant at the first glance into a single
proposition.

Case 1. $S=\mathbf{skip}$. Obvious.

Case 2. $S=q:=0$. We only consider the case of
$type(q)=\mathbf{integer}$, and the case of
$type(q)=\mathbf{Boolean}$ is similar. First, it holds that
\begin{equation*}\begin{split}tr((\sum_{n=-\infty}^{\infty}|n\rangle_q\langle
0|P|0\rangle_q\langle n|)\rho)&
=tr(P\sum_{n=-\infty}^{\infty}|0\rangle_q\langle n|\rho
|n\rangle_q\langle 0|)\\ &=tr(P[|q:=0|](\rho)).
\end{split}\end{equation*} On the other hand, for any quantum predicate $Q\in\mathcal{P}(\mathcal{H}_{all})$, if $\models_{{\rm
tot}}\{Q\}q:=0\{P\}$, i.e. \begin{equation*}\begin{split}
tr(Q\rho)&\leq tr(P[|q:=0|](\rho))\\
&=tr((\sum_{n=-\infty}^{\infty}|n\rangle_q\langle
0|P|0\rangle_q\langle n|)\rho)
\end{split}\end{equation*} for all
$\rho\in\mathcal{D}^{-}(\mathcal{H}_{{\rm all}})$, then it follows
from Lemma~\ref{order} that $$Q\sqsubseteq
\sum_{n=-\infty}^{\infty}|n\rangle_q\langle 0|P|0\rangle_q\langle
n|.$$

Case 3. $S=\overline{q}:=U\overline{q}$. Similar to Case 2.

Case 4. $S=S_1;S_2$. It follows from the induction hypothesis on
$S_1$ and $S_2$ that
\begin{equation*}\begin{split}
tr((wp.S_1.(wp.S_2.P))\rho)&=tr((wp.S_2.P)[|S_1|](\rho))\\
&=tr(P[|S_2|]([|S_1|](\rho)))\\
&=tr(P[|S_1;S_2|](\rho)).
\end{split}\end{equation*}If $\models_{{\rm
tot}}\{Q\}S_1;S_2\{P\}$, then for all
$\rho\in\mathcal{D}^{-}(\mathcal{H}_{{\rm all}})$, we have:
$$tr(QP)\leq tr(P[|S_1;S_2|](\rho))=tr((wp.S_1.(wp.S_2.P))\rho).$$ Therefore, it follows from
Lemma~\ref{order} that $Q\sqsubseteq wp.S_1.(wp.S_2.P)$.

Case 5. $S=\mathbf{measure}\ M[\overline{q}]:\overline{S}$. Applying
the induction hypothesis on $S_m$, we obtain:
\begin{equation*}\begin{split}
tr((\sum_m M_m^{\dag}(wp.S_m.P)M_m)\rho)&=\sum_mtr((wp.S_m.P)M_m\rho
M_m^{\dag})\\ &=\sum_m tr(P[|S_m|](M_m\rho
M_m^{\dag}))\\ &=tr(P\sum_m [|S_m|](M_m\rho M_m^{\dag}))\\
&=tr(P[|\mathbf{measure}\ M[\overline{q}]:\overline{S}|]_{{\rm
par}}(\rho)).
\end{split}\end{equation*}

If $\models_{{\rm tot}}\{Q\}\mathbf{measure}\
M[\overline{q}]:\overline{S}\{P\}$, then $$tr(Q\rho)\leq tr((\sum_m
M_m^{\dag}(wp.S_m.P)M_m)\rho)$$ for all $\rho$, and it follows from
Lemma~\ref{order} that $$Q\sqsubseteq \sum_m
M_m^{\dag}(wlp.S_m.P)M_m.$$

Case 6. $S=\mathbf{while}\ M[\overline{q}]=1\ \mathbf{do}\
S^{\prime}$. For simplicity, we write $(\mathbf{while})^{n}$ for
statement \textquotedblleft$(\mathbf{while}\ M[\overline{q}]=1\
\mathbf{do}\ S^{\prime})^{n}$\textquotedblright. First, we have:
$$tr(P_n\rho)=tr(P[|(\mathbf{while})^{n}|](\rho)).$$ This
claim can be proved by induction on $n$. The basis case of $n=0$ is
obvious. By the induction hypotheses on $n$ and $S^{\prime}$, we
obtain:
\begin{equation*}\begin{split}tr(P_{n+1}\rho)&=tr(M_0^{\dag}PM_0\rho)+tr(M_1^{\dag}(wp.S^{\prime}.P_n)M_1\rho)\\
&= tr(PM_0\rho M_0^{\dag})+tr((wp.S^{\prime}.P_n)M_1\rho M_1^{\dag})\\
&= tr(PM_0\rho M_0^{\dag})+tr(P_n[|S^{\prime}|](M_1\rho
M_1^{\dag}))\\ &= tr(PM_0\rho
M_0^{\dag})+tr(P[|(\mathbf{while})^{n}|]([|S^{\prime}|]_{{\rm
par}}(M_1\rho M_1^{\dag})))\\
&=tr[P(M_0\rho M_0^{\dag}+[|S^{\prime};(\mathbf{while})^{n}|](M_1\rho M_1^{\dag}))]\\
&=tr(P[|(\mathbf{while})^{n+1}|](\rho)).
\end{split}\end{equation*}

Now continuity of trace operator yields:
\begin{equation*}\begin{split}
tr((\bigvee_{n=0}^{\infty}P_n)\rho)&=\bigvee_{n=0}^{\infty}
tr(P_n\rho)\\
&=\bigvee_{n=0}^{\infty}tr(P[|(\mathbf{while})^{n}|](\rho))\\
&=tr(P\bigvee_{n=0}^{\infty}[|(\mathbf{while})^{n}|](\rho))\\
&=tr(P[|\mathbf{while}\ M[\overline{q}]=1\ \mathbf{do}\
S^{\prime}|](\rho)).
\end{split}\end{equation*}So, if $$\models_{{\rm tot}}\{Q\}\mathbf{while}\ M[\overline{q}]=1\
\mathbf{do}\ S^{\prime}\{P\},$$ then $$tr(Q\rho)\leq
tr((\bigvee_{n=0}^{\infty}P_n)\rho)$$ for all $\rho$, and by
Lemma~\ref{order} we obtain $Q\sqsubseteq
\bigvee_{n=0}^{\infty}P_n$. $\Box$

\begin{prop}\label{wlp-re}\begin{enumerate}\item \begin{enumerate}\item $wlp.\mathbf{skip}.P=P.$
\item If $type(q)=\mathbf{Boolean}$, then $$wlp.q:=0.P=|0\rangle_q\langle
0|P|0\rangle_q\langle 0|+|1\rangle_q\langle 0|P|0\rangle_q\langle
1|,$$ and if $type(q)=\mathbf{integer}$, then
$$wlp.q:=0.P=\sum_{n=-\infty}^{\infty} |n\rangle_q\langle 0|P|0\rangle_q\langle
n|.$$ \item $wlp.\overline{q}:=U\overline{q}.P=U^{\dag}PU$.
\item $wlp.S_1;S_2.P=wlp.S_1.(wlp.S_2.P)$. \item $wlp.\mathbf{measure}\
M[\overline{q}]:\overline{S}.P=\sum_m M_m^{\dag}(wlp.S_m.P)M_m.$
\item $wlp.\mathbf{while}\ M[\overline{q}]=1\ \mathbf{do}\
S.P=\bigwedge_{n=0}^{\infty}P_n,$ where $$\begin{cases}&P_0=I_{\mathcal{H}_{all}},\\
& P_{n+1}=M_0^{\dag}PM_0+M_1^{\dag}(wlp.S.P_n)M_1\ {\rm for\ all}\
n\geq 0.
\end{cases}$$
\end{enumerate}\item For any quantum program $S$, for any quantum
predicate $P\in\mathcal{P}(\mathcal{H}_{{\rm all}}),$ and for any
partial density operator $\rho\in\mathcal{D}^{-}(\mathcal{H}_{{\rm
all}})$, we have: $$tr((wlp.S.P)\rho)=tr(P[|S|]_{{\rm
par}}(\rho))+[tr(\rho)-tr([|S|](\rho)].$$\end{enumerate}
\end{prop}

\textit{Proof.} Similar to the case of weakest precondition, we
prove (1) and (2) simultaneously by induction on the structure of
quantum program $S$.

Case 1. $S=\mathbf{skip}$, or $q:=0$, or
$\overline{q}:=U\overline{q}$. Similar to Cases 1, 2 and 3 in the
proof of Proposition~\ref{wp-re}.

Case 2. $S=S_1;S_2$. First, with the induction hypothesis on $S_1$
and $S_2$, we have:
\begin{equation*}\begin{split}&tr(wlp.S_1.(wlp.S_2.P)\rho)=tr(wlp.S_2.P[|S_1|](\rho))+[tr(\rho)-tr([|S_1|](\rho))]\\
&=tr(P[|S_2|]([|S_1|](\rho))+[tr([|S_1|](\rho))-tr([|S_2|]([|S_1|](\rho)))]
+[tr(\rho)-tr([|S_1|](\rho))]\\
&=tr(P[|S_2|]([|S_1|](\rho))+[tr(\rho)-tr([|S_2|]([|S_1|](\rho)))]\\
&=tr(P[|S|](\rho))+[tr(\rho)-tr([|S|](\rho))].
\end{split}\end{equation*}If $\models_{\rm par}\{Q\}S\{P\}$, then it
holds that $$tr(Q\rho)\leq
tr(P[|S|](\rho))+[tr(\rho)-tr([|S|](\rho))]=tr(wlp.S_1.(wlp.S_2.P)\rho)$$
for all $\rho\in\mathcal{D}^{-}(\mathcal{H}_{all})$, and by
Lemma~\ref{order} we obtain: $$Q\sqsubseteq wlp.S_1.(wlp.S_2.P).$$

Case 3. $S=\mathbf{measure}\ M[\overline{q}]:\overline{S}$. It can
is derived by induction hypothesis on all $S_m$ that
\begin{equation*}\begin{split}&tr(\sum_m
M_m^{\dag}(wlp.S_m.P)M_m\rho)=\sum_m
tr(M_m^{\dag}(wlp.S_m.P)M_m\rho)\\ &=\sum_m tr((wlp.S_m.P)M_m\rho
M_m^{\dag})\\ &=\sum_m\{tr(P[|S_m|](M_m\rho M_m^{\dag}))+[tr(M_m\rho
M_m^{\dag})-tr([|S_m|](M_m\rho M_m^{\dag}))]\}\\
&=\sum_m tr(P[|S_m|](M_m\rho M_m^{\dag}))+[\sum_m tr(M_m\rho
M_m^{\dag})-\sum_m tr([|S_m|](M_m\rho M_m^{\dag}))]\}\\
&=tr(P\sum_m[|S_m|](M_m\rho M_m^{\dag}))+[tr(\rho
\sum_m M_m^{\dag}M_m)-tr(\sum_m[|S_m|](M_m\rho M_m^{\dag}))]\}\\
&=tr(P[|S|](\rho))+[tr(\rho)-tr([|S|](\rho))]
\end{split}\end{equation*} because
$\sum_mM_m^{\dag}M_m=I_{\mathcal{H}_{\overline{q}}}.$ If
$\models_{\rm par}\{Q\}S\{P\}$, then for all
$\rho\in\mathcal{D}^{-}(\mathcal{H}_{all})$, it holds that
$$tr(Q\rho)\leq tr(P[|S|](\rho))+[tr(\rho)-tr([|S|](\rho))]=tr(\sum_m
M_m^{\dag}(wlp.S_m.P)M_m\rho).$$ This together with
Lemma~\ref{order} implies $$Q\sqsubseteq \sum_m
M_m^{\dag}(wlp.S_m.P)M_m.$$

Case 4. $S=\mathbf{while}\ M[\overline{q}]=1\ \mathbf{do}\
S^{\prime}$. We first prove that
\begin{equation}\label{wlp-med}tr(P_n\rho)=tr(P[|(\mathbf{while})^{n}|](\rho))+
[tr(\rho)-tr([|(\mathbf{while})^{n}|](\rho))]\end{equation} by
induction on $n$, where $(\mathbf{while})^{n}$ is an abbreviation of
$(\mathbf{while}\ M[\overline{q}]=1\ \mathbf{do}\ S^{\prime})^{n}$.
The case of $n=0$ is obvious. By induction on $S^{\prime}$ and
induction hypothesis on $n$, we observe:
\begin{equation*}\begin{split}
&tr(P_{n+1}\rho)=tr[(M_0^{\dag}PM_0)+M_1^{\dag}(wlp.S^{\prime}.P_n)M_1\rho]\\
&=tr(M_0^{\dag}PM_0\rho)+tr(M_1^{\dag}(wlp.S^{\prime}.P_n)M_1\rho)\\
&=tr(PM_0\rho M_0^{\dag})+tr((wlp.S^{\prime}.P_n)M_1\rho
M_1^{\dag})\\ &=tr(PM_0\rho M_0^{\dag})+tr(P_n[|S^{\prime}|](M_1\rho
M_1^{\dag}))+[tr(M_1\rho M_1^{\dag})-tr([|S^{\prime}|](M_1\rho
M_1^{\dag}))]\\
&=tr(PM_0\rho M_0^{\dag})+tr(P[|(\mathbf{while})^{n}|]([|S|](M_1\rho
M_1^{\dag})))+[tr([|S|](M_1\rho M_1^{\dag}))\\
&\ \ \ \ \ \ \ \ \ \ -tr([|(\mathbf{while})^{n}|]([|S|](M_1\rho
M_1^{\dag})))]+[tr(M_1\rho M_1^{\dag})-tr([|S^{\prime}|](M_1\rho
M_1^{\dag}))]\\
&=tr(P[M_0\rho M_0^{\dag}+[|(\mathbf{while})^{n}|]([|S|](M_1\rho
M_1^{\dag}))]\\ &\ \ \ \ \ \ \ \ \ \ +[tr(\rho)-tr(M_0\rho
M_0^{\dag}+[|(\mathbf{while})^{n}|]([|S|](M_1\rho M_1^{\dag})))]\\
&=tr(P[|(\mathbf{while})^{n+1}|](\rho))+[tr(\rho)-tr([|(\mathbf{while})^{n+1}|](\rho)].
\end{split}\end{equation*}This completes the proof of
Eq.~(\ref{wlp-med}). Note that quantum predicate $P\sqsubseteq I$.
Then $I-P$ is positive, and by continuity of trace operator we
obtain:\begin{equation*}\begin{split}&tr((\bigwedge_{n=0}^{\infty}P_n)\rho)=\bigwedge_{n=0}^{\infty}tr(P_n\rho)\\
&=\bigwedge_{n=0}^{\infty}\{tr(P[|(\mathbf{while})^{n}|](\rho))+[tr(\rho)-tr([|(\mathbf{while})^{n}|](\rho))]\}\\
&=tr(\rho)+\bigwedge_{n=0}^{\infty}tr[(P-I)[|(\mathbf{while})^{n}|](\rho)]\\
&=tr(\rho)+tr[(P-I)\bigvee_{n=0}^{\infty}[|(\mathbf{while})^{n}|](\rho)]\\
&=tr(\rho)+tr[(P-I)[|S|](\rho)]\\
&=tr(P[|S|](\rho))+[tr(\rho)-tr([|S|](\rho))].
\end{split}\end{equation*}For any
$Q\in\mathcal{P}(\mathcal{H}_{all})$, $\models_{\rm par}\{Q\}S\{P\}$
implies: $$tr(Q\rho)\leq
tr(P[|S|](\rho))+[tr(\rho)-tr([|S|](\rho))=tr((\bigwedge_{n=0}^{\infty}P_n)\rho)$$
for all $\rho\in\mathcal{D}^{-}(\mathcal{H}_{all})$. This together
with Lemma~\ref{order} leads to
$Q\sqsubseteq\bigwedge_{n=0}^{\infty}P_n$. $\Box$

We see that Propositions~\ref{wp-re} and~\ref{wlp-re} coincide with
Figs. 2 and 3, respectively, in~\cite{FDJY07} when only quantum
variables of type \textbf{Boolean} are considered. But Figs. 2 and 3
in~\cite{FDJY07} are given directly as definitions, and their
intuitive meanings are not clear because they do not have any
connection to total and partial correctness of quantum programs. In
contrast, Propositions~\ref{wp-re} and~\ref{wlp-re} fill in such a
gap, and they are derived from Definition~\ref{wlp-def}, which is
given entirely based on the notions of total and partial correctness
of quantum programs.

The next proposition gives a recursive characterization of weakest
precondition and weakest liberal precondition of quantum loop, and
it provides a key step in the proof of completeness of quantum Hoare
logic.

\begin{prop}\label{wlp-recur}We write $\mathbf{while}$ for quantum loop \textquotedblleft$\mathbf{while}\
M[\overline{q}]=1\ \mathbf{do}\ S$\textquotedblright. Then for any
$P\in\mathcal{P}(\mathcal{H}_{all})$, we have:\begin{enumerate}\item
$wp.\mathbf{while}.P=M_0^{\dag}PM_0+M_1^{\dag}(wp.S.(wp.\mathbf{while}.P))M_1.$
\item
$wlp.\mathbf{while}.P=M_0^{\dag}PM_0+M_1^{\dag}(wlp.S.(wlp.\mathbf{while}.P))M_1.$
\end{enumerate}
\end{prop}

\textit{Proof.} We only prove (2), and the proof of (1) is similar
and easier. For every $\rho\in\mathcal{D}^{-}(\mathcal{H}_{all})$,
by Proposition~\ref{wlp-re}(2) we observe:
\begin{equation*}\begin{split}&tr[(M_0^{\dag}PM_0+M_1^{\dag}(wlp.S.(wlp.\mathbf{while}.P))M_1)\rho]\\
&=tr(PM_0\rho M_0^{\dag})+tr[(wlp.S.(wlp.\mathbf{while}.P))M_1\rho
M_1^{\dag}]\\
&=tr(PM_0\rho M_0^{\dag})+tr[(wlp.\mathbf{while}.P)[|S|](M_1\rho
M_1^{\dag})]\\ &\ \ \ \ \ \ \ \ \ \ \ \ \ \ \ \ \ \ \ \ \ \ \ \ \ \
\ \ \ \ \ \ \ \ \ \ \ \ \ \ +[tr(M_1\rho
M_1^{\dag})-tr([|S|](M_1\rho
M_1^{\dag}))]\\
&=tr(PM_0\rho M_0^{\dag})+tr[P[|\mathbf{while}|]([|S|](M_1\rho
M_1^{\dag}))]+[tr([|S|](M_1\rho M_1^{\dag})\\
&\ \ \ \ \ \ \ \ \ \ -tr([|\mathbf{while}|]([|S|](M_1\rho
M_1^{\dag}))] +[tr(M_1\rho M_1^{\dag})-tr([|S|](M_1\rho
M_1^{\dag}))]\\
&=tr[P(M_0\rho M_0^{\dag}+[|\mathbf{while}|]([|S|](M_1\rho
M_1^{\dag})))]\\ &\ \ \ \ \ \ \ \ \ \ \ \ \ \ \ \ \ \ \ \ \ \ \ \ \
\ \ \ \ \ \ \ +[tr(M_1\rho
M_1^{\dag})-tr([|\mathbf{while}|]([|S|](M_1\rho M_1^{\dag}))]\\
&=tr(P[|\mathbf{while}|](\rho))+[tr(\rho
M_1^{\dag}M_1)-tr([|\mathbf{while}|]([|S|](M_1\rho M_1^{\dag})))]\\
&=tr(P[|\mathbf{while}|](\rho))+[tr(\rho (I-
M_0^{\dag}M_0))-tr([|\mathbf{while}|]([|S|](M_1\rho M_1^{\dag})))]\\
&=tr(P[|\mathbf{while}|](\rho))+[tr(\rho)-tr(
M_0\rho M_0^{\dag}+[|\mathbf{while}|]([|S|](M_1\rho M_1^{\dag})))]\\
&=tr(P[|\mathbf{while}|](\rho))+[tr(\rho)-tr([|\mathbf{while}|](\rho))].
\end{split}\end{equation*}This means that $$\{M_0^{\dag}PM_0+M_1^{\dag}
(wlp.S.(wlp.\mathbf{while}.P))M_1\}\mathbf{while}\{P\},$$ and
$$Q\sqsubseteq
M_0^{\dag}PM_0+M_1^{\dag}(wlp.S.(wlp.\mathbf{while}.P))M_1$$
provided $\models_{{\rm par}}\{Q\}\mathbf{while}\{P\}$. $\Box$

\section{Proof System for Partial Correctness}

Now we are ready to present an axiomatic system of Hoare logic for
quantum programs. The quantum Hoare logic can be divided into two
proof systems, one for partial correctness and one for total
correctness. In this section, we introduce the proof system $qPD$
for partial correctness of quantum programs. It consists of the
axioms and inference rules in Fig. 2.

\begin{figure}[h]\centering
\begin{equation*}\begin{split}
&(Axiom\ Skip)\ \ \ \ \ \ \ \ \ \ \ \ \ \ \ \ \ \ \ \ \ \ \ \ \ \ \ \ \{P\}\mathbf{Skip}\{P\}\\
& \\
&(Axiom\ Initialization)\ {\rm If}\ type(q)=\mathbf{Boolean},\ {\rm then}\\
& \\ &\ \ \ \ \ \ \ \ \ \ \ \ \ \ \ \ \ \ \ \ \ \ \ \ \ \ \ \ \
\{|0\rangle_q\langle 0|P|0\rangle_q\langle 0|+|1\rangle_q\langle
0|P|0\rangle_q\langle 1|\}q:=0\{P\}
\\ & \\ & {\rm
and\ if}\ type(q)=\mathbf{integer},\ {\rm then}\\ & \\ &\ \ \ \ \ \
\ \ \ \ \ \ \ \ \ \ \ \ \ \ \
 \ \ \ \ \ \ \ \ \ \ \ \ \ \ \ \ \ \ \ \ \{\sum_{n=-\infty}^{\infty}|n\rangle_q\langle 0|P|0\rangle_q\langle
n|\}q:=0\{P\}\\
& \\ &(Axiom\ Unitary\ Transformation)\ \ \ \ \ \ \ \ \ \
\{U^{\dag}PU\}\overline{q}:=U\overline{q}\{P\}\\
& \\ &(Rule\ Sequential\ Composition)\ \ \ \ \ \ \ \ \ \
\frac{\{P\}S_1\{Q\}\ \ \ \ \ \ \{Q\}S_2\{R\}}{\{P\}S_1;S_2\{R\}}\\
&\\
&(Rule\ Measurement)\ \ \ \ \ \ \ \ \ \ \ \ \ \
\frac{\{P_m\}S_m\{Q\}\ {\rm for\ all}\ m}{\{\sum_m
M_m^{\dag}P_mM_m\}\mathbf{measure}\
M[\overline{q}]:\overline{S}\{Q\}}\\
&\\
&(Rule\ Loop\ Partial)\ \ \ \ \ \ \ \ \ \ \ \ \ \
\frac{\{Q\}S\{M_0^{\dag}PM_0+M_1^{\dag}QM_1\}}{\{M_0^{\dag}PM_0+M_1^{\dag}QM_1\}\mathbf{while}\
M[\overline{q}]=1\ \mathbf{do}\ S\{P\}}\\
&\\
&(Rule\ Order)\ \ \ \ \ \ \ \ \ \ \ \ \ \ \frac{P\sqsubseteq
P^{\prime}\ \ \ \ \{P^{\prime}\}S\{Q^{\prime}\}\ \ \ \
Q^{\prime}\sqsubseteq Q}{\{P\}S\{Q\}}
\end{split}\end{equation*}
\caption{Proof System $qPD$ of Partial Correctness}\label{fig 2}
\end{figure}

We first prove the soundness of the proof system $qPD$ with respect
to the semantics of partial correctness: provability of a
correctness formula in $qPD$ implies its truth in the sense of
partial correctness.

\begin{thm} (Soundness) The proof system $qPD$ is sound for partial correctness of quantum programs; that is,
for any quantum program $S$ and
quantum predicates $P,Q\in\mathcal{P}(\mathcal{H}_{{\rm all}})$, we
have:
$$\vdash_{qPD}\{P\}S\{Q\}\ {\rm implies}\ \models_{{\rm par}}\{P\}S\{Q\}.$$
\end{thm}

\textit{Proof.} We only need to show that the axioms and inference
rules of $qPD$ are valid in the sense of partial correctness.

(Axiom Skip) It is obvious that $\models_{{\rm
par}}\{P\}\mathbf{skip}\{P\}.$

(Axiom Initialization) We only prove the case of
$type(q)=\mathbf{integer}$, and the case of
$type(q)=\mathbf{Boolean}$ is similar. For any
$\rho\in\mathcal{D}^{-}(\mathcal{H}_{{\rm all}})$, it follows from
Proposition~\ref{sempro}.2 that \begin{equation*}\begin{split}
tr[(\sum_{n=-\infty}^{\infty}|n\rangle_q\langle
0|P|0\rangle_q\langle
n|)\rho]&=\sum_{n=-\infty}^{\infty}tr(|n\rangle_q\langle
0|P|0\rangle_q\langle n|\rho)\\
&=\sum_{n=-\infty}^{\infty}tr(P|0\rangle_q\langle
n|\rho|n\rangle_q\langle
0|)\\
&=tr(P\sum_{n=-\infty}^{\infty}|0\rangle_q\langle
n|\rho|n\rangle_q\langle
0|)\\
&=tr(P[|q:=0|](\rho)).
\end{split}\end{equation*} Therefore, we have: $$\models_{{\rm par}}\{\sum_{n=-\infty}^{\infty}|n\rangle_q\langle
0|P|0\rangle_q\langle n|\}q:=0\{P\}.$$

(Axiom Unitary Transformation) It is easy to see that
$$\models_{{\rm par}}\{U^{\dag}PU\}\overline{q}:=U\overline{q}\{P\}.$$

(Rule Sequential Composition) If $\models_{{\rm par}}\{P\}S_1\{Q\}$
and $\models_{{\rm par}}\{Q\}S_2\{R\}$, then for any
$\rho\in\mathcal{D}^{-}(\mathcal{H}_{all})$ we have:
\begin{equation*}\begin{split}tr(P\rho)&\leq
tr(Q[|S_1|](\rho))+[tr(\rho)-tr([|S_1|](\rho))]\\
&\leq
tr(R[|S_2|]([|S_1|](\rho)))+[tr([|S_1|](\rho))-tr([|S_2|]([|S_1|](\rho)))]\\ &\ \ \ \ \ \ \ \ \ \ \ \
\ \ \ \ \ \ \ \ \ \ \ \ \ \ \ \ \ \ \ \ \ \ \ \ \ \ \ \ \ \ \ \ \ \ \ \ +[tr(\rho)-tr([|S_1|](\rho))]\\
&=tr(R[|S_1;S_2|](\rho))+[tr(\rho)-tr([|S_1;S_2|](\rho))].
\end{split}\end{equation*}
Therefore, $\models_{{\rm par}}\{P\}S_1;S_2\{R\}$ as desired.

(Rule Measurement) Assume that $\models_{par}\{P_m\}S_m\{Q\}$ for
all $m$. Then for all $\rho\in\mathcal{D}^{-}(\mathcal{H}_{all})$,
since $\sum_mM_m^{\dag}M_m=I_{\mathcal{H}_{\overline{q}}}$, it holds
that
\begin{equation*}\begin{split}&tr(\sum_mM_m^{\dag}P_mM_m\rho)=\sum_mtr(M_m^{\dag}P_mM_m\rho)\\
&=\sum_mtr(P_mM_m\rho M_m^{\dag})\\
&\leq\sum_m\{tr(Q[|S_m|](M_m\rho M_m^{\dag}))+[tr(M_m\rho
M_m^{\dag})-tr([|S_m|](M_m \rho M_m^{\dag}))]\}\\
&\leq \sum_m tr(Q[|S_m|](M_m\rho M_m^{\dag}))+[\sum_m tr(M_m\rho
M_m^{\dag})-\sum_m tr([|S_m|](M_m \rho M_m^{\dag}))]\}\\
&=tr(Q\sum_m [|S_m|](M_m\rho M_m^{\dag}))+[tr(\sum_m \rho
M_m^{\dag}M_m)- tr(\sum_m [|S_m|](M_m \rho M_m^{\dag}))]\}\\
&=tr(Q[|\mathbf{measure}|](\rho))+[tr(\rho)-tr([|\mathbf{measure}|](\rho)],
\end{split}\end{equation*} and $$\models_{{\rm par}}\{\sum_m
M_m^{\dag}P_mM_m\}\mathbf{measure}\{Q\},$$ where $\mathbf{measure}$
is an abbreviation of statement \textquotedblleft$\mathbf{measure}\
M[\overline{q}]:\overline{S}$\textquotedblright.

(Rule Loop Partial) Suppose that $$\models_{{\rm
par}}\{Q\}S\{M_0^{\dag}PM_0+M_1^{\dag}QM_1.$$ Then for all
$\rho\in\mathcal{D}^{-}(\mathcal{H}_{all})$, it holds that
\begin{equation}\label{sou-med0}tr(Q\rho)\leq
tr((M_0^{\dag}PM_0+M_1^{\dag}QM_1)[|S|](\rho))+[tr(\rho)-tr([|S|](\rho))].\end{equation}
Furthermore, we have:
\begin{equation}\label{sou-med}\begin{split} tr[(M_0^{\dag}PM_0+M_1^{\dag}QM_1)\rho]&\leq
\sum_{k=0}^{n}tr(P(\mathcal{E}_0\circ
([|S|]\circ\mathcal{E}_1)^{k})(\rho))\\ &\ \ \ \ \ \
+tr(Q(\mathcal{E}_1\circ ([|S|]\circ\mathcal{E}_1)^{n})(\rho))\\ &\
\ \ \ \ \ +\sum_{k=0}^{n-1}[tr(\mathcal{E}_1\circ ([|S|]\circ
\mathcal{E}_1)^{k}(\rho))-tr(([|S|]\circ\mathcal{E}_1)^{k+1}(\rho))]\end{split}
\end{equation} for all $n\geq 1$.
In fact, Eq.~(\ref{sou-med}) may be proved by induction on $n$. The
case of $n=1$ is obvious. Using Eq.~(\ref{sou-med0}), we obtain:
\begin{equation}\label{sou-med1}\begin{split}
&tr(Q(\mathcal{E}_1\circ ([|S|]\circ\mathcal{E}_1)^{n})(\rho))\leq
tr((M_0^{\dag}PM_0+M_1^{\dag}QM_1)([|S|]\circ\mathcal{E}_1)^{n+1}(\rho))\\
&\ \ \ \ \ \ \ \ \ \ \ \ \ \ \ \ \ \ \ \ \ \ \ \ \ \ \ \ \ \ \ \ \ \
\ \ \ \ \ \ \ \ \ \ \ \ \ \ \ \ +
[tr((\mathcal{E}_1\circ([|S|]\circ\mathcal{E}_1)^{n})(\rho))-tr(([|S|]\circ\mathcal{E}_1)^{n+1}(\rho))]\\
&=tr(P(\mathcal{E}_0\circ([|S|]\circ\mathcal{E}_1)^{n+1})(\rho))+
tr(Q(\mathcal{E}_1\circ([|S|]\circ\mathcal{E}_1)^{n+1})(\rho))\\ &\
\ \ \ \ \ \ \ +
[tr((\mathcal{E}_1\circ([|S|]\circ\mathcal{E}_1)^{n})(\rho))-tr(([|S|]\circ\mathcal{E}_1)^{n+1}(\rho))].
\end{split}\end{equation}Combining Eqs.~(\ref{sou-med}) and
(\ref{sou-med1}), we assert that
\begin{equation*}\begin{split}tr[(M_0^{\dag}PM_0+M_1^{\dag}QM_1)\rho]&\leq
\sum_{k=0}^{n+1}tr(P(\mathcal{E}_0\circ
([|S|]\circ\mathcal{E}_1)^{k})(\rho))\\ &\ \ \ \ \ \
+tr(Q(\mathcal{E}_1\circ ([|S|]\circ\mathcal{E}_1)^{n+1})(\rho))\\
&\ \ \ \ \ \
 +\sum_{k=0}^{n}[tr(\mathcal{E}_1\circ ([|S|]\circ
\mathcal{E}_1)^{k}(\rho))-tr(([|S|]\circ\mathcal{E}_1)^{k+1}(\rho))].\end{split}
\end{equation*}Therefore, Eq.~(\ref{sou-med}) holds in the case of
$n+1$ provided it is true in the case of $n$, and we complete the
proof of Eq.~(\ref{sou-med}).

Now we note that \begin{equation*}\begin{split}
tr((\mathcal{E}_1\circ([|S|]\circ\mathcal{E}_1)^{k}(\rho))&=tr(M_1([|S|]\circ\mathcal{E}_1)^{k}(\rho)M_1^{\dag})\\
&=tr(([|S|]\circ\mathcal{E}_1)^{k}(\rho)M_1^{\dag}M_1)\\
&=tr(([|S|]\circ\mathcal{E}_1)^{k}(\rho)(I-M_0^{\dag}M_0))\\
&=tr(([|S|]\circ\mathcal{E}_1)^{k}(\rho))-tr((\mathcal{E}_0\circ([|S|]\circ\mathcal{E}_1)^{k})(\rho)).
\end{split}\end{equation*}
Then it follows that \begin{equation}\label{sou-med2}\begin{split}
&\sum_{k=0}^{n-1}[tr(\mathcal{E}_1\circ ([|S|]\circ
\mathcal{E}_1)^{k}(\rho))-tr(([|S|]\circ\mathcal{E}_1)^{k+1}(\rho))]=
\sum_{k=0}^{n-1}tr(([|S|]\circ \mathcal{E}_1)^{k}(\rho))\\ &\ \ \ \
\ \ \ \ \ \ \ \ \ \ \ \ \ \ \ \ \
-\sum_{k=0}^{n-1}[tr(\mathcal{E}_0\circ ([|S|]\circ
\mathcal{E}_1)^{k}(\rho)) -\sum_{k=0}^{n-1}tr(([|S|]\circ
\mathcal{E}_1)^{k+1}(\rho))\\
&=tr(\rho)-tr(([|S|]\circ\mathcal{E}_1)^{n}(\rho))-\sum_{k=0}^{n-1}tr(\mathcal{E}_0\circ
([|S|]\circ \mathcal{E}_1)^{k}(\rho)).
\end{split}\end{equation}On the other hand, we have:
\begin{equation}\label{sou-med3}\begin{split}
tr(Q(\mathcal{E}_1\circ([|S|]\circ\mathcal{E}_1)^{n})(\rho))&=
tr(QM_1([|S|]\circ\mathcal{E}_1)^{n})(\rho)M_1^{\dag})\\ &\leq tr(M_1([|S|]\circ\mathcal{E}_1)^{n})(\rho)M_1^{\dag})\\
&= tr(([|S|]\circ\mathcal{E}_1)^{n})(\rho)M_1^{\dag}M_1)\\
&= tr(([|S|]\circ\mathcal{E}_1)^{n})(\rho)(I-M_0^{\dag}M_0))\\
&=
tr(([|S|]\circ\mathcal{E}_1)^{n})(\rho))-tr((\mathcal{E}_0\circ([|S|]\circ\mathcal{E}_1)^{n})(\rho)).
\end{split}\end{equation}Putting Eqs.~(\ref{sou-med2}) and
(\ref{sou-med3}) into Eq.~(\ref{sou-med}), we obtain:
\begin{equation*}\begin{split}
tr[(M_0^{\dag}PM_0+M_1^{\dag}QM_1)\rho]& \leq
\sum_{k=0}^{n}tr(P(\mathcal{E}_0\circ
([|S|]\circ\mathcal{E}_1)^{k})(\rho))\\
&\ \ \ \ +[tr(\rho)-\sum_{k=0}^{n}tr((\mathcal{E}_0\circ
([|S|]\circ\mathcal{E}_1)^{k})(\rho))]\\
&=tr(P\sum_{k=0}^{n}(\mathcal{E}_0\circ
([|S|]\circ\mathcal{E}_1)^{k})(\rho))\\
&\ \ \ \ +[tr(\rho)-tr(\sum_{k=0}^{n}(\mathcal{E}_0\circ
([|S|]\circ\mathcal{E}_1)^{k})(\rho))].
\end{split}\end{equation*}Let $n\rightarrow\infty$. Then it follows
that $$tr[(M_0^{\dag}PM_0+M_1^{\dag}QM_1)\rho]\leq
tr(P[|\mathbf{while}|](\rho)+[tr(\rho)-tr([|\mathbf{while}|](\rho))]$$
and
$$\models_{{\rm par}}\{M_0^{\dag}PM_0+M_1^{\dag}QM_1\}\mathbf{while}\{P\},$$
where $\mathbf{while}$ is an abbreviation of quantum loop
\textquotedblleft$\mathbf{while}\ M[\overline{q}]=1\ \mathbf{do}\
S$\textquotedblright. $\Box$

Now we are going to establish completeness for the proof system
$qPD$ with respect to the semantics of partial correctness: truth of
a quantum program in the sense of partial correctness implies its
provability in $qPD$. Note that the L$\ddot{o}$wner ordering
assertions between quantum predicates in (Rule Order) are statements
about complex numbers. So, only a completeness of $qPD$ relative to
the theory of the field of complex numbers may be anticipated; more
precisely, we can add all statements that are true in the field of
complex numbers into $qPD$ in order to make it complete. The
following theorem should be understood exactly in the sense of such
a relative completeness.

\begin{thm}\label{par-comp} (Completeness) The proof system $qPD$ is complete for partial
correctness of quantum programs;
that is, for any quantum program $S$ and quantum predicates
$P,Q\in\mathcal{P}(\mathcal{H}_{{\rm all}})$, we have:
$$\models_{{\rm par}}\{P\}S\{Q\}\ {\rm implies}\ \vdash_{qPD}\{P\}S\{Q\}.$$
\end{thm}

\textit{Proof.} If $\models_{{\rm par}}\{P\}S\{Q\}$, then by
Definition~\ref{wlp-def} (2) we have $P\sqsubseteq wlp.S.Q$.
Therefore, by (Rule Order) it suffices to prove the following:
\begin{itemize}\item\ Claim: $\vdash_{qPD} \{wlp.S.Q\}S\{Q\}.$\end{itemize}

We proceed by induction on the structure of $S$.

Case 1. $S=\mathbf{skip}$. Immediate from (Axiom Skip).

Case 2. $S=q:=0$. Immediate from (Axiom Initialization).

Case 3. $S=\overline{q}:=U\overline{q}$. Immediate from (Axiom
Unitary Transformation).

Case 4. $S=S_1;S_2$. It follows from the induction hypothesis on
$S_1$ and $S_2$ that
$$\vdash_{qPD}\{wlp.S_1.(wlp.S_2.Q)\}S_1\{wlp.S_2.Q\}$$ and $$\vdash_{qPD}\{wlp.S_2.Q\}S_2\{Q\}.$$
We obtain: $$\vdash_{qPD}\{wlp.S_1.(wlp.S_2.Q)\}S_1;S_2\{Q\}$$ by
(Rule Sequential Composition). Then with Proposition~\ref{wlp-re}
(1.d) we see that $$\vdash_{qPD}\{wlp.S_1;S_2.Q\}S_1;S_2\{Q\}.$$

Case 5. $S=\mathbf{measure}\ M[\overline{q}]:\overline{S}$. For all
$m$, by the induction hypothesis on $S_m$ we obtain:
$$\vdash_{qPD}\{wlp.S_m.Q\}S_m\{Q\}.$$ Then applying (Rule
Measurement) yields: $$\vdash_{qPD}\{\sum_m
M_m^{\dag}(wlp.S_m.Q)M_m\}\mathbf{measure}\
M[\overline{q}]:\overline{S}\{Q\},$$ and using
Proposition~\ref{wlp-re} (1.e) we have:
$$\vdash_{qPD}\{wlp.\mathbf{measure}\
M[\overline{q}]:\overline{S}.P\}\mathbf{measure}\
M[\overline{q}]:\overline{S}\{Q\}.$$

Case 6. $S=\mathbf{while}\ M[\overline{q}]=1\ \mathbf{do}\
S^{\prime}$. For simplicity, we write $\mathbf{while}$ for quantum
loop \textquotedblleft$\mathbf{while}\ M[\overline{q}]=1\
\mathbf{do}\ S^{\prime}$\textquotedblright. The induction hypothesis
on $S$ asserts that
$$\vdash_{qPD}\{wlp.S.(wlp.\mathbf{while}.P)\}S\{wlp.\mathbf{while}.P\}.$$ By
Proposition~\ref{wlp-recur}(2) we have:
$$wlp.\mathbf{while}.P=M_0^{\dag}PM_0+M_1^{\dag}(wlp.S.(wlp.\mathbf{while}.P))M_1.$$
Then by (Rule Loop Par) we obtain:
$$\vdash_{qPD}\{wlp.\mathbf{while}.P\}\mathbf{while}\{P\}$$ as
desired. $\Box$

\section{Proof System for Total Correctness}

The aim of this section is to present a proof system $qTD$ for
correctness of quantum program. The only difference between $qTD$
and $qPD$ is the inference rule for quantum loops. To give the rule
for total correctness of quantum loops, we need a notion of bound
function which express the number of iterations of a quantum loop in
its computation.

\begin{defn}\label{bou-def}Let $P\in\mathcal{P}(\mathcal{H}_{{\rm all}})$ and
$\epsilon>0$. A function $t: \mathcal{D}^{-}(\mathcal{H}_{{\rm
all}})\rightarrow\mathbb{N}$ is called a $(P,\epsilon)-$bound
function of quantum loop \textquotedblleft$\mathbf{while}\
M[\overline{q}]=1\ \mathbf{do}\ S$\textquotedblright if it satisfies
the following conditions:\begin{enumerate}\item $t([|S|](M_1\rho
M_1^{\dag}))\leq t(\rho)$; and \item $tr(P\rho)\geq\epsilon$ implies
$t([|S|](M_1\rho M_1^{\dag}))< t(\rho)$\end{enumerate} for all
$\rho\in \mathcal{D}^{-}(\mathcal{H}_{{\rm all}})$.
\end{defn}

Recall that a bound function $t$ of a classical loop
\textquotedblleft$\mathbf{while}\ B\ \mathbf{do}\ S\ \mathbf{od}$
\textquotedblright satisfies the inequality $t([|S|](s))<t(s)$ for
any input state $s$. It is interesting to compare it with conditions
(1) and (2) of the above definition, and we see that the latter are
two inequalities between $t([|S|](M_1\rho M_1^{\dag}))$ and
$t(\rho)$ but not between $t([|S|](\rho))$ and $t(\rho)$. This is
because in the implementation of the quantum loop
\textquotedblleft$\mathbf{while}\ M[\overline{q}]=1\ \mathbf{do}\
S$\textquotedblright, we need to perform the yes-no measurement $M$
on $\rho$ when checking the loop guard
\textquotedblleft$M[\overline{q}]=1$\textquotedblright, and the
state of quantum variables will become $M_1\rho M_1^{\dag}$ from
$\rho$ whence the measurement outcome \textquotedblleft
yes\textquotedblright is observed.

The following lemma gives a characterization of the existence of
bound function of a quantum loop in terms of the limit of the state
of quantum variables when the number of iterations of the loop goes
to infinity. It provides a key step for the proof of soundness and
completeness of the proof system $qTD$.

\begin{lem}\label{bou-lem}Let $P\in\mathcal{P}(\mathcal{H}_{{\rm
all}})$. Then the following two statements are equivalent:
\begin{enumerate}\item for any $\epsilon>0$, there exists a $(P,\epsilon)-$bound
function $t_\epsilon$ of quantum loop
\textquotedblleft$\mathbf{while}\ M[\overline{q}]=1\ \mathbf{do}\
S$\textquotedblright;\item
$\lim_{n\rightarrow\infty}tr(P([|S|]\circ\mathcal{E}_1)^{n}(\rho))=0$
for all $\rho\in\mathcal{D}^{-}(\mathcal{H}_{{\rm all}})$.
\end{enumerate}
\end{lem}

\textit{Proof.} (1 $\Rightarrow$ 2) We prove this by refutation. If
$$\lim_{n\rightarrow\infty}tr(P([|S|]\circ\mathcal{E}_1)^{n}(\rho))\neq
0,$$ then there exist $\epsilon_0>0$ and strictly increasing
sequence $\{n_k\}$ of nonnegative integers such that
$$tr(P([|S|]\circ\mathcal{E}_1)^{n_k}(\rho))\geq\epsilon_0$$ for all
$k\geq 0$. Thus, we have a $(P,\epsilon_0)-$bound function of loop
$\mathbf{while}\ M[\overline{q}]=1\ \mathbf{do}\ S$. For each $k\geq
0$, we set $$\rho_k=([|S|]\circ\mathcal{E}_1)^{n_k}(\rho).$$ Then it
holds that $tr(P\rho_k)\geq\epsilon_0$, and by conditions 1 and 2 in
Definition~\ref{bou-def} we obtain:\begin{equation*}\begin{split}t
t_{\epsilon_0}(\rho_k)&>t_{\epsilon_0}([|S|](M_1\rho_k
M_1^{\dag}))\\
&=t_{\epsilon_0}(([|S|]\circ\mathcal{E}_1)(\rho_k))\\
&\geq
t_{\epsilon_0}(([|S|]\circ\mathcal{E}_1)^{n_{k+1}-n_k}(\rho_k))\\
&=t_{\epsilon_0}(\rho_{k+1}).
\end{split}\end{equation*}
Consequently, we have an infinitely descending chain
$\{t_{\epsilon_0}(\rho_k)\}$ in $\mathbb{N}$, the set of nonnegative
integers. This is a contradiction because $\mathbb{N}$ is a
well-founded set.

(2 $\Rightarrow$ 1) For each
$\rho\in\mathcal{D}^{-}(\mathcal{H}_{{\rm all}})$, if
$$\lim_{n\rightarrow\infty}tr(P([|S|]\circ\mathcal{E}_1)^{n}(\rho))=0,$$
then for any $\epsilon>0$, there exists $N\in\mathbb{N}$ such that
$$tr(P([|S|]\circ\mathcal{E}_1)^{n}(\rho))<\epsilon$$ for all $n\geq
N$. We define:
$$t_\epsilon(\rho)=\min\{N\in\mathbb{N}:tr(P([|S|]\circ\mathcal{E}_1)^{n}(\rho))<\epsilon\ {\rm for\ all}\ n\geq N\}.$$
Now it suffices to show that $t_\epsilon$ is a $(P,\epsilon)-$bound
function of loop \textquotedblleft$\mathbf{while}\
M[\overline{q}]=1\ \mathbf{do}\ S$\textquotedblright. To this end.
we consider the following two cases:

Case 1. $tr(P\rho)\geq\epsilon$. Suppose that $t_\epsilon(\rho)=N$.
Then $tr(P\rho)\geq\epsilon$ implies $N\geq 1$. By the definition of
$t_\epsilon$, we assert that
$$tr(P([|S|]\circ\mathcal{E}_1)^{n}(\rho))<\epsilon$$ for all $n\geq
N$. Thus, for all $n\geq N-1\geq 0$,
$$tr(P([|S|]\circ\mathcal{E}_1)^{n}([|S|](M_1^{\dag}\rho
M_1)))=tr(P([|S|]\circ\mathcal{E}_1)^{n+1}(\rho))<\epsilon.$$
Therefore, $$t_\epsilon([|S|](M_1^{\dag}\rho M_1))\leq
N-1<N=t_\epsilon(\rho).$$

Case 2. $tr(P\rho)<\epsilon$. Again, suppose that
$t_\epsilon(\rho)=N$. Now we have the following two subcases:

Subcase 2.1. $N=0$. Then for all $n\geq 0$,
$$tr(P([|S|]\circ\mathcal{E}_1)^{n}(\rho))<\epsilon.$$ It is easy to
see that $t_\epsilon([|S|](M_1\rho M_1^{\dag}))=0=t_\epsilon(\rho).$

Subcase 2.2. $n\geq 1$. We can derive that
$$t_\epsilon(\rho)>t_\epsilon([|S|](M_1\rho M_1^{\dag}))$$ in the way
of Case 1. $\Box$

Now we are ready to present the proof system $qTD$ for total
correctness of quantum programs. It consists of the axioms (Axiom
Skip), (Axiom Initialization) and (Axiom Unitary Transformation) and
inference rules (Rule Sequential Composition), (Rule Measurement)
and (Rule Order) in Fig.2 as well as inference rule (Rule Loop
Total) in Fig.3.

\begin{figure}[h]\centering
\begin{equation*}\begin{split}
&(Rule\ Loop\ Total)\ \ \ \ \ \ \ \ \ \ \ \ \ \
\frac{\begin{split}&\{Q\}S\{M_0^{\dag}PM_0+M_1^{\dag}QM_1\}\\{\rm
for\ any}\ \epsilon>0,\ &t_\epsilon\ {\rm is\ a}\
(M_1^{\dag}QM_1,\epsilon)-{\rm bound\ function}\\ &{\rm of\ loop}\
\mathbf{while}\ M[\overline{q}]=1\ \mathbf{do}\
S\end{split}}{\{M_0^{\dag}PM_0+M_1^{\dag}QM_1\}\mathbf{while}\
M[\overline{q}]=1\ \mathbf{do}\ S\{P\}}
\end{split}\end{equation*}
\caption{Proof System $qTD$ of Total Correctness}\label{fig 3}
\end{figure}

The remainder of this section will be devoted to establish soundness
and completeness of $qTD$: provability of a correctness formula in
$qTD$ is equivalent to its truth in the sense of total correctness.

\begin{thm} (Soundness) The proof system $TD$ is sound for total correctness of quantum programs; that is, for any quantum program $S$ and
quantum predicates $P,Q\in\mathcal{P}(\mathcal{H}_{{\rm all}})$, we
have:
$$\vdash_{qTD}\{P\}S\{Q\}\ {\rm implies}\ \models_{{\rm tot}}\{P\}S\{Q\}.$$
\end{thm}

\textit{Proof.} It suffices to show that the axioms and inference
rules of $TD$ are valid in the sense of total correctness.

The proof for soundness of (Axiom Skip), (Axiom Initialization) and
(Axiom Unitary Transformation) is similar to the case of partial
correctness.

(Rule Sequential Composition) Suppose that $\models_{{\rm
tot}}\{P\}S_1\{Q\}$ and $\models_{{\rm tot}}\{Q\}S_2\{R\}$. Then for
any $\rho\in\mathcal{D}^{-}(\mathcal{H}_{{\rm all}})$, with
Proposition~\ref{sempro}.4 we obtain:
\begin{equation*}\begin{split}tr(P\rho)&\leq tr(Q[|S_1|](\rho))\\
&\leq tr(R[|S_2|]([|S_1|](\rho)))\\ &=tr(P[|S_1;S_2|](\rho)).
\end{split}\end{equation*}Therefore, $\models_{{\rm
tot}}\{P\}S_1;S_2\{R\}.$

(Rule Measurement) Suppose that $\models_{{\rm tot}}\{P_m\}S_m\{Q\}$
for all $m$. Then for any $\rho\in\mathcal{D}^{-}(\mathcal{H}_{{\rm
all}})$, it holds that
$$tr(P_mM_m\rho M_m^{\dag})\leq tr(Q[|S_m|](M_m\rho
M_m^{\dag}))$$ because $\{P_m\}S_m\{Q\}$ for all $m$. Therefore, we
have: \begin{equation*}\begin{split} tr(\sum_m
M_m^{\dag}P_mM_m\rho)&=\sum_m tr(P_mM_m\rho M_m^{\dag})\\
&\leq\sum_m tr(Q[|S_m|](M_m\rho M_m^{\dag}))\\
&=tr(Q\sum_m [|S_m|](M_m\rho M_m^{\dag}))\\
&=tr(Q[|\mathbf{measure}\ M[\overline{q}]:\overline{S}|](\rho)),
\end{split}\end{equation*} and $$\models_{{\rm tot}}\{\sum_mM_m^{\dag}PM_m\}\mathbf{measure}\ M[\overline{q}]:\overline{S}
\{Q\}.$$

(Rule Loop Total) If
$$\models_{{\rm tot}}\{Q\}S\{M_0^{\dag}PM_0+M_1^{\dag}QM_1\},$$ then for any
$\rho\in\mathcal{D}^{-}(\mathcal{H}_{{\rm all}})$, we have:
\begin{equation}\label{loop1}tr(Q\rho)\leq tr((M_0^{\dag}PM_0+M_1^{\dag}QM_1)[|S|]_{{\rm
par}}(\rho)).\end{equation}

We first prove the following inequality:
\begin{equation}\label{ineq}\begin{split}tr[(M_0^{\dag}PM_0&+M_1^{\dag}QM_1)\rho]\\ &\leq
\sum_{k=0}^{n}tr(P[\mathcal{E}_0\circ ([|S|]\circ
\mathcal{E}_1)]^{k}(\rho))+tr(Q[\mathcal{E}_1\circ
([|S|]\circ\mathcal{E}_1)^{n}](\rho))\end{split}\end{equation} by
induction on $n$. It holds that
\begin{equation*}\begin{split}tr[(M_0^{\dag}PM_0+M_1^{\dag}QM_1)\rho]&=tr(PM_0\rho
M_0^{\dag})+tr(QM_1\rho M_1^{\dag})\\
&=tr(P\mathcal{E}_0(\rho))+tr(Q\mathcal{E}_1(\rho)).\end{split}\end{equation*}
So, Eq.~(\ref{ineq}) is correct for the base case of $n=0$. Assume
Eq.~(\ref{ineq}) is correct for the case of $n=m$. Then applying
Eq.~(\ref{loop1}), we obtain:
\begin{equation*}\begin{split}
&tr[(M_0^{\dag}PM_0+M_1^{\dag}QM_1)\rho]=tr(P\mathcal{E}_0(\rho))+tr(QM_1\rho M_1^{\dag})\\
&\leq \sum_{k=0}^{m}tr(P[\mathcal{E}_0\circ ([|S|]\circ
\mathcal{E}_1)]^{k}(\rho))+tr(Q[\mathcal{E}_1\circ
([|S|]\circ\mathcal{E}_1)^{m}](\rho))\\
&\leq \sum_{k=0}^{m}tr(P[\mathcal{E}_0\circ ([|S|]\circ
\mathcal{E}_1)]^{k}(\rho))+tr((M_0^{\dag}PM_0+M_1^{\dag}QM_1)
[|S|]([\mathcal{E}_1\circ
([|S|]\circ\mathcal{E}_1)^{m}](\rho)))\\
&= \sum_{k=0}^{m}tr(P[\mathcal{E}_0\circ ([|S|]\circ
\mathcal{E}_1)]^{k}(\rho))+ tr(PM_0[|S|]([\mathcal{E}_1\circ
([|S|]\circ\mathcal{E}_1)^{m}](\rho))M_0^{\dag})\\ &\ \ \ \ \ \ \ \
\ \ \ \ \ \ \ \ \ \ \ \ \ \ \ \ \ \ \ \ \ \ \ \ \ \ \ \ \ \ \ \ \ \
\ \ \ \ \ \ \ \ \ \ + tr(QM_1[|S|]([\mathcal{E}_1\circ
([|S|]\circ\mathcal{E}_1)^{m}](\rho))M_1^{\dag})\\
&=\sum_{k=0}^{m+1}tr(P[\mathcal{E}_0\circ ([|S|]\circ
\mathcal{E}_1)]^{k}(\rho))+tr(Q[\mathcal{E}_1\circ
([|S|]\circ\mathcal{E}_1)^{m+1}](\rho)).
\end{split}\end{equation*}Therefore, Eq.~(\ref{ineq}) also holds for
the case of $n=M+1$. Now, since for any $\epsilon>0$, there exists
$M_1^{\dag}QM_1,\epsilon-$bound function $t_\epsilon$ of quantum
loop \textquotedblleft$\mathbf{while}\ M[\overline{q}]=1\
\mathbf{do}\ S$\textquotedblright, by Lemma~\ref{bou-lem} we have:
\begin{equation*}\begin{split}\lim_{n\rightarrow\infty}tr(Q[\mathcal{E}_1\circ([|S|]\circ\mathcal{E}_1)^{n}(\rho))&=
\lim_{n\rightarrow\infty}tr(QM_1([|S|]\circ\mathcal{E}_1)^{n}(\rho)M_1^{\dag})\\&=\lim_{n\rightarrow\infty}tr(M_1^{\dag}QM_1
([|S|]\circ\mathcal{E}_1)^{n}(\rho))=0.
\end{split}\end{equation*}Consequently, it holds that
\begin{equation*}\begin{split}
tr[(M_0^{\dag}PM_0+M_1^{\dag}QM_1)\rho]& \leq
\lim_{n\rightarrow\infty}\sum_{k=0}^{n}tr(P[\mathcal{E}_0\circ
([|S|]\circ \mathcal{E}_1)]^{n}(\rho))\\
&\ \ \ \ \ \ \ \ \ \
+\lim_{n\rightarrow\infty}tr(Q[\mathcal{E}_1\circ
([|S|]\circ\mathcal{E}_1)^{n}](\rho))\\ & =
\sum_{n=0}^{\infty}tr(P[\mathcal{E}_0\circ ([|S|]\circ
\mathcal{E}_1)]^{n}(\rho))\\
&=tr(P\sum_{n=0}^{\infty}[\mathcal{E}_0\circ ([|S|]\circ \mathcal{E}_1)^{n}](\rho))\\
&=tr(P[|\mathbf{while}\ M[\overline{q}]=1\ \mathbf{do}\ S|](\rho)).\
\Box
\end{split}\end{equation*}

\begin{thm} (Completeness) The proof system $TD$ is complete for total correctness of quantum programs; that is, for any quantum program $S$ and
quantum predicates $P,Q\in\mathcal{P}(\mathcal{H}_{{\rm all}})$, we
have:
$$\models_{{\rm tot}}\{P\}S\{Q\}\ {\rm implies}\ \vdash_{qTD}\{P\}S\{Q\}.$$
\end{thm}

\textit{Proof.} Similar to the case of partial correctness, it
suffices to prove the following:
\begin{itemize}\item \ Claim: $\vdash_{qTD} \{wlp.S.Q\}S\{Q\}$ for any quantum program $S$
and quantum predicate $P\in\mathcal{P}(\mathcal{H}_{{\rm
all}})$\end{itemize} because by Definition~\ref{wlp-def} (1) we have
$P\sqsubseteq wp.S.Q$ when $\models_{tot}\{P\}S\{Q\}.$ The above
claim can be done by induction on the structure of $S$. We only
consider the case of $S=\mathbf{while}\ M[\overline{q}]=1\
\mathbf{do}\ S^{\prime}$, and the other cases are similar to the
proof of Theorem~\ref{par-comp}. We write $\mathbf{while}$ for
quantum loop \textquotedblleft$\mathbf{while}\ M[\overline{q}]=1\
\mathbf{do}\ S^{\prime}$\textquotedblright. It follows from
Proposition~\ref{wlp-recur}(1) that
$$wp.\mathbf{while}.Q=M_0^{\dag}QM_0+M_1^{\dag}(wp.S.(wp.\mathbf{while}.Q))M_1.$$
So, our aim is to derive that
$$\vdash_{qTD}\{M_0^{\dag}QM_0+M_1^{\dag}(wp.S.(wp.\mathbf{while}.Q))M_1\}\mathbf{while}\{Q\}.$$
By the induction hypothesis on $S^{\prime}$ we get:
$$\vdash_{qTD}\{wp.S^{\prime}.(wp.\mathbf{while}.Q)\}S\{wp.\mathbf{while}.Q\}.$$
Then by (Rule Loop Total) it suffices to show that for any
$\epsilon>0$, there exists a $(M_1^{\dag}(wp.S^{\prime}.$
$(wp.S.Q))M_1,\epsilon)-$bound function of quantum loop
$\mathbf{while}$. Applying Lemma~\ref{bou-lem}, we only need to
prove:
\begin{equation}\label{lim-0}\lim_{n\rightarrow\infty}tr(M_1^{\dag}(wp.S^{\prime}.(wp.\mathbf{while}.Q))
M_1([|S^{\prime}|]\circ \mathcal{E}_1)^{n}(\rho))=0.\end{equation}
First, by Propositions~\ref{wp-re} (2) and \ref{sempro} (6) we
observe:\begin{equation}\label{remain}\begin{split}
tr(M_1^{\dag}(wp.S^{\prime}.&(wp.\mathbf{while}.Q))
M_1([|S^{\prime}|]\circ \mathcal{E}_1)^{n}(\rho))\\
&=tr(wp.S^{\prime}.(wp.\mathbf{while}.Q)
M_1([|S^{\prime}|]\circ \mathcal{E}_1)^{n}(\rho)M_1^{\dag})\\
&=tr(wp.\mathbf{while}.Q[|S^{\prime}|](M_1([|S^{\prime}|]\circ
\mathcal{E}_1)^{n}(\rho)M_1^{\dag}))\\
&=tr(wp.\mathbf{while}.Q([|S^{\prime}|]\circ
\mathcal{E}_1)^{n+1}(\rho))\\
&=tr(Q[|\mathbf{while}|]([|S^{\prime}|]\circ
\mathcal{E}_1)^{n+1}(\rho))\\
&=\sum_{k=n+1}^{\infty}tr(Q[\mathcal{E}_0\circ ([|S^{\prime}|]\circ
\mathcal{E}_1)^{k}](\rho)).
\end{split}\end{equation}Second, we consider the following infinite
series of nonnegative real numbers:
\begin{equation}\label{series}\begin{split}\sum_{n=0}^{\infty}tr(Q[\mathcal{E}_0\circ
([|S^{\prime}|]\circ\mathcal{E}_1)^{k}](\rho))
=tr(Q\sum_{n=0}^{\infty}[\mathcal{E}_0\circ
([|S^{\prime}|]\circ\mathcal{E}_1)^{k}](\rho)).\end{split}\end{equation}
Since $Q\sqsubseteq I_{\mathcal{H}_{all}}$, it follows from
Propositions~\ref{sempro} (6) and~\ref{sempro1} that
\begin{equation*}\begin{split}tr(Q\sum_{n=0}^{\infty}[\mathcal{E}_0\circ
([|S^{\prime}|]\circ\mathcal{E}_1)^{k}](\rho))
&=tr(Q[|\mathbf{while}|](\rho))\\ &\leq
tr([|\mathbf{while}|](\rho))\leq tr(\rho)\leq
1.\end{split}\end{equation*}Therefore, the infinite series
Eq.~(\ref{series}) converges. Note that Eq.~(\ref{remain}) is the
sum of the remaining terms of the infinite series Eq.~(\ref{series})
after the $n$th term. Then convergence of the infinite series
Eq.~(\ref{series}) implies Eq.~(\ref{lim-0}), and we complete the
proof. $\Box$

It should be pointed out that the above theorem is merely a relative
completeness of $qTD$ with respect to the theory of the fields of
complex numbers because except that (Rule Order) is employed in
$qTD$, the existence of bound functions in (Rule Loop Total) is also
a statement about complex numbers.

\section{Conclusion}

Based on D'Hondt and Panangaden's idea of representing quantum
predicates by Hermitian operators and Selinger's idea of modeling
quantum programs by super-operators, a full-fledged Hoare logic is
developed for deterministic quantum programs, and its completeness
relative to the theory of the field of complex numbers is proved in
this paper.

An interesting problem for further studies is to find reasonable
extensions of the quantum Hoare logic presented in this paper for
bigger classes of quantum programs, including nondeterministic
quantum programs~\cite{Zu04}, parallel and distributed quantum
programs. Hopefully, they will serve as a logical foundation of
various effective techniques for verification of these bigger
classes of quantum programs.

\newpage

\section*{Appendix: Proof of Proposition~\ref{change}}

(1) We proceed by induction on the structure of $S$.

Case 1. $S=\mathbf{skip}$. Obvious.

Case 2. $S=q:=0$. We only consider the case of
$type(q)=\mathbf{integer}$, and the case of
$type(q)=\mathbf{Boolean}$ is similar. It holds that
\begin{equation*}\begin{split}
tr_{var(S)}([|S|](\rho))&=\sum_{n,n^{\prime}=-\infty}^{\infty}
\mathbf{}_q\langle n^{\prime}|0\rangle_q\langle
n|\rho|n\rangle_q\langle 0|n^{\prime}\rangle_q\\
&=\sum_{n=-\infty}^{\infty} \mathbf{}_q\langle n|\rho|n\rangle_q\\
&=tr_{var(S)}(\rho).
\end{split}\end{equation*}

Case 3. $S=\overline{q}:=U\overline{q}$. If $\{|\psi_i\rangle\}$ is
an orthonormal basis of $\mathcal{H}_{\overline{q}}$, then
$\{U^{\dag}|\psi_i\rangle\}$ is also an orthonormal basis of
$\mathcal{H}_{\overline{q}}$. Consequently,
$$tr_{Var(S)}([|S|](\rho))=\sum_i \langle\psi_i|U\rho U^{\dag}|\psi_i\rangle=tr_{Var(S)}(\rho).$$

Case 4. $S=S_1;S_2$. By the induction hypothesis on $S_1$ and $S_2$,
we have:
\begin{equation*}\begin{split}tr_{var(S)}([|S|](\rho))&=tr_{var(S_1)\cup
var(S_2)}([|S_2|]([|S_1|](\rho)))\\
&=tr_{var(S_1)}(tr_{var(S_2)}([|S_2|]([|S_1|](\rho))))\\
&=tr_{var(S_1)}(tr_{var(S_2)}([|S_1|](\rho)))\\
&=tr_{var(S_2)}(tr_{var(S_1)}([|S_1|](\rho)))\\
&=tr_{var(S_2)}(tr_{var(S_1)}(\rho))\\ &=tr_{var(S)}(\rho).
\end{split}\end{equation*}

Case 5. $S=\mathbf{measure}\ M[\overline{q}]:\overline{S}$. By the
induction hypothesis on $S_m$ for all outcome $m$ of measurement
$M$, we
obtain:\begin{equation*}\begin{split}tr_{var(S)}([|S|](\rho))&=\sum_m
tr_{var(S)}([|S_m|](M_m\rho M_m^{\dag}))\\ &=\sum_m
tr_{var(S)-var(S_m)}(tr_{var(S_m)}([|S_m|](M_m\rho M_m^{\dag})))\\
&=\sum_m tr_{var(S)-var(S_m)}(tr_{var(S_m)}(M_m\rho M_m^{\dag}))\\
&=\sum_m tr_{var(S)}(M_m\rho M_m^{\dag})\\
&=\sum_m tr_{var(S)-\{\overline{q}\}}(tr_{\overline{q}}(M_m\rho
M_m^{\dag}))\\&=\sum_m
tr_{var(S)-\{\overline{q}\}}(tr_{\overline{q}}(M_m^{\dag}M_m\rho))
\\&=tr_{var(S)-\{\overline{q}\}}(tr_{\overline{q}}((\sum_m M_m^{\dag}M_m)\rho))
\\&=tr_{var(S)-\{\overline{q}\}}(tr_{\overline{q}}(\rho))\\
&=tr_{var(S)}(\rho)
\end{split}
\end{equation*} because $\sum_m
M_m^{\dag}M_m=I_{\mathcal{H}_{\overline{q}}}$.

Case 6. $S=\mathbf{while}\ M[\overline{q}]=1\ \mathbf{do}\
S^{\prime}$. For simplicity, we write $\mathbf{while}$ for quantum
loop \textquotedblleft$\mathbf{while}\ M[\overline{q}]=1\
\mathbf{do}\ S^{\prime}$\textquotedblright. First, we have:
$$tr_{var(S)}([|(\mathbf{while})^{0}|](\rho))=0_{\mathcal{H}_{all}},$$
\begin{equation}\label{var}tr_{var(S)}([|(\mathbf{while})^{n}|](\rho))=tr_{var(S)}(\rho)\end{equation}
for all $n\geq 1$. This can be proved by induction on $n$. In fact,
$$(\mathbf{while})^{k+1}=\mathbf{measure}\
M[\overline{q}]:\overline{S}$$ where $\overline{S}=S_0,S_1$,
$S_0=\mathbf{skip}$, and $S_1=S^{\prime};(\mathbf{while})^{k}$.
Thus, by Cases 1, 4 and 5, we can derive Eq.~(\ref{var}) for $n=k+1$
immediately from the induction hypothesis for $n=k$. Now, it follows
from continuity of trace that
\begin{equation*}\begin{split}tr_{var(S)}([|S|](\rho))&=tr_{var(S)}(\bigvee_{n=0}^{\infty}[|(\mathbf{while})^{n}|](\rho))\\
&=\bigvee_{n=0}^{\infty}tr_{var(S)}([|(\mathbf{while})^{n}|](\rho))\\
&=tr_{var(S)}(\rho).
\end{split}\end{equation*}

(2) The trick is to prove the following slightly stronger
conclusion:
\begin{itemize}\item \ Claim: For any $var(S)\subseteq X\subseteq Var$,
$tr_{X-var(S)}(\rho_1)=tr_{X-var(S)}(\rho_2)$ implies
$tr_{X-var(S)}([|S|](\rho_1))=tr_{X-var(S)}([|S|](\rho_2)).$
\end{itemize}

We also proceed by induction on the structure of $S$.

Case 1. $S=\mathbf{skip}.$ Obvious.

Case 2. $S=q:=0$. We only consider the case of
$type(q)=\mathbf{integer}$, the case of $type(q)=\mathbf{Boolean}$
is similar. First, let $\{|\psi_i\rangle\}$ be an orthonormal basis
of $\mathcal{H}_{X-\{q\}}$, then we
have:\begin{equation*}\begin{split}tr_{X-var(S)}([|S|](\rho))&=tr_{X-\{q\}}
[\sum_{n=-\infty}^{\infty}(|0\rangle_q\langle n|\otimes
I_{\mathcal{H}_{Var-\{q\}}})\rho(|n\rangle_q\langle
0|\otimes I_{\mathcal{H}_{Var-\{q\}}})]\\
&=\sum_i\sum_{n=-\infty}^{\infty}(I_{\mathcal{H}_{(Var-X)\cup\{q\}}}\otimes\langle\psi_i|)(|0\rangle_q\langle
n|\otimes I_{\mathcal{H}_{Var-\{q\}}})\rho\\
&\ \ \ \ \ \ \ \ \ \ \ \ \ \ \ \ \ \ \ \ \ \ \ \ \ \ \ \ \ \
(|n\rangle_q\langle 0|\otimes
I_{\mathcal{H}_{Var-\{q\}}})(I_{\mathcal{H}_{(Var-X)\cup\{q\}}}\otimes
|\psi_i\rangle)\\
&=\sum_{n=-\infty}^{\infty}((|0\rangle_q\langle n|\otimes
I_{\mathcal{H}_{Var-\{q\}}})\sum_i(I_{\mathcal{H}_{(Var-X)\cup\{q\}}}\otimes\langle\psi_i|)\rho\\
&\ \ \ \ \ \ \ \ \ \ \ \ \ \ \ \ \ \ \ \ \ \ \ \ \ \ \ \ \ \
(I_{\mathcal{H}_{(Var-X)\cup\{q\}}}\otimes
|\psi_i\rangle)(|n\rangle_q\langle 0|\otimes
I_{\mathcal{H}_{Var-\{q\}}})\\
&=\sum_{n=-\infty}^{\infty}|0\rangle_q\langle
n|tr_{X-var(S)}(\rho)|n\rangle_q\langle 0|.
\end{split}\end{equation*}Thus, it is easy to see that (2) holds in this case.

Case 3. $S=\overline{q}:=U\overline{q}$. Assume that
$\{|\psi_i\rangle\}$ is an orthonormal basis of
$\mathcal{H}_{X-\{\overline{q}\}}$. Then it holds that
\begin{equation*}\begin{split}&tr_{X-var(S)}([|S|](\rho))=tr_{X-\{\overline{q}\}}(U\otimes
I_{\mathcal{H}_{Var-\{\overline{q}\}}})\rho (U^{\dag}\otimes
I_{\mathcal{H}_{Var-\{\overline{q}\}}})\\
&=\sum_i(I_{\mathcal{H}_{(Var-X)\cup\{\overline{q}\}}}\otimes\langle\psi_i|)(U\otimes
I_{\mathcal{H}_{Var-\{\overline{q}\}}})\rho (U^{\dag}\otimes
I_{\mathcal{H}_{Var-\{\overline{q}\}}})(I_{\mathcal{H}_{(Var-X)\cup\{\overline{q}\}}}\otimes
|\psi_i\rangle)\\
&=(U\otimes
I_{\mathcal{H}_{Var-\{\overline{q}\}}})\sum_i(I_{\mathcal{H}_{(Var-X)\cup\{\overline{q}\}}}\otimes\langle\psi_i|)\rho
(I_{\mathcal{H}_{(Var-X)\cup\{\overline{q}\}}}\otimes
|\psi_i\rangle)(U^{\dag}\otimes
I_{\mathcal{H}_{Var-\{\overline{q}\}}})\\
&=(U\otimes
I_{\mathcal{H}_{Var-\{\overline{q}\}}})tr_{X-var(S)}(\rho)(U^{\dag}\otimes
I_{\mathcal{H}_{Var-\{\overline{q}\}}})
\end{split}\end{equation*}because
$|\psi_i\rangle\in\mathcal{H}_{X-\{\overline{q}\}}$,
$U\in\mathcal{L}(\mathcal{H}_{\overline{q}})$, and
$(X-\{\overline{q}\})\cap\{\overline{q}\}=\emptyset$. Now it is easy
to see that (2) holds in this case.

Case 4. $S=S_1;S_2$. If $var(S_1)\cup var(S_2)=var(S)\subseteq X$,
then there are $Y,Z\subseteq Var$ such that $var(S_1)\subseteq Y$,
$var(S_2)\subseteq Z$ and $$Y-var(S_1)=X-var(S)=Z-var(S_2).$$ Thus,
we have:
\begin{equation*}\begin{split}tr_{Y-var(S_1)}(\rho_1)&=tr_{X-var(S)}(\rho_1)\\
&=tr_{X-var(S)}(\rho_2)\\
&=tr_{Y-var(S_1)}(\rho_2).
\end{split}\end{equation*}
By the induction hypothesis on $S_1$, we obtain: \begin{equation*}
\begin{split}tr_{Z-var(S_2)}([|S_1|](\rho_1))&=tr_{Y-var(S_1)}([|S_1|](\rho_1))\\
&=tr_{Y-var(S_1)}([|S_1|](\rho_2))\\
&=tr_{Z-var(S_2)}([|S_1|](\rho_2)).
\end{split}\end{equation*}Furthermore, by the induction hypothesis on $S_2$,
we obtain:\begin{equation*}\begin{split}tr_{X-var(S)}([|S|](\rho_1))&=tr_{Z-var(S_2)}([|S_2|]([|S_1|](\rho_1)))\\
&=tr_{Z-var(S_2)}([|S_2|]([|S_1|](\rho_2)))\\
&=tr_{(X-var(S)}([|S|](\rho_2)).\end{split}\end{equation*}

Case 5. $S=\mathbf{measure}\ M[\overline{q}]:\overline{S}$. First,
let $\{|\psi_i\rangle\}$ be an orthonormal basis of
$\mathcal{H}_{X-var(S)}$. Then for any outcome $m$ of measurement
$M$, we have:
\begin{equation*}\begin{split}tr_{X-var(S)}(M_m\rho
M_m^{\dag})&=\sum_i(I_{\mathcal{H}_{Var-(X-var(S))}}\otimes\langle\psi_i|)(M_m\otimes
I_{\mathcal{H}_{Var-\{\overline{q}\}}})\rho\\ &\ \ \ \ \ \ \ \ \ \ \
\ \ \ \ \ (M_m^{\dag}\otimes I_{\mathcal{H}_{Var-\{\overline{q}\}}})
(I_{\mathcal{H}_{Var-(X-var(S))}}\otimes
|\psi_i\rangle)\\
&=(M_m\otimes
I_{\mathcal{H}_{Var-\{\overline{q}\}}})\sum_i(I_{\mathcal{H}_{Var-(X-var(S))}}\otimes\langle\psi_i|)\rho\\
&\ \ \ \ \ \ \ \ \ \ \ \ \ \ \ \
(I_{\mathcal{H}_{Var-(X-var(S))}}\otimes
|\psi_i\rangle)(M_m^{\dag}\otimes
I_{\mathcal{H}_{Var-\{\overline{q}\}}})\\
&=M_mtr_{X-var(S)}(\rho)M_m^{\dag}
\end{split}\end{equation*}because
$|\psi_i\rangle\in\mathcal{H}_{X-var(S)}$,
$M_m\in\mathcal{L}(\mathcal{H}_{\overline{q}})$ and
$(X-var(S))\cap\{\overline{q}\}=\emptyset$. Consequently, it follows
from $tr_{X-var(S)}(\rho_1)=tr_{X-var(S)}(\rho_2)$ that
$$tr_{X-var(S)}(M_m\rho_1M_m^{\dag})=tr_{X-var(S)}(M_m\rho_2M_m^{\dag}).$$
Note that we can write $X-var(S)=Y-var(S_m)$ for some $Y\supseteq
var(S_m)$. Then it holds that
$$tr_{Y-var(S_m)}(M_m\rho_1M_m^{\dag})=tr_{Y-var(S_m)}(M_m\rho_2M_m^{\dag}).$$
By the induction hypothesis on $S_m$, we obtain:
\begin{equation*}\begin{split}tr_{X-var(S)}([|S_m|](M_m\rho_1M_m^{\dag}))&=tr_{Y-var(S_m)}([|S_m|](M_m\rho_1M_m^{\dag}))\\
&=tr_{Y-var(S_m)}([|S_m|](M_m\rho_2M_m^{\dag}))\\
&=tr_{X-var(S)}([|S_m|](M_m\rho_2M_m^{\dag})).\end{split}\end{equation*}
Therefore, it follows that \begin{equation*}\begin{split}
tr_{X-var(S)}([|S|](\rho_1))&=tr_{X-var(S)}\sum_m[|S_m|](M_m\rho_1M_m^{\dag})\\
&=\sum_mtr_{X-var(S)}([|S_m|](M_m\rho_1M_m^{\dag}))\\
&=\sum_mtr_{X-var(S)}([|S_m|](M_m\rho_2M_m^{\dag}))\\
&=tr_{X-var(S)}\sum_m[|S_m|](M_m\rho_2M_m^{\dag})\\
&=tr_{X-var(S)}([|S|](\rho_2)).
\end{split}\end{equation*}

Case 6. $S=\mathbf{while}\ M[\overline{q}]=1\ \mathbf{do}\
S^{\prime}$. Assume that
$tr_{X-var(S)}(\rho_1)=tr_{X-var(S)}(\rho_2)$. Using Cases 1, 4 and
5, we can show that
$$tr_{X-var(S)}([|(\mathbf{while})^{n}|](\rho_1))=tr_{X-var(S)}([|(\mathbf{while})^{n}|](\rho_2))$$
for all $n\geq 0$ by induction on $n$. Therefore, we obtain:
\begin{equation*}\begin{split}
tr_{X-var(S)}([|\mathbf{while}|](\rho_1))&=tr_{X-var(S)}(\bigvee_{n=0}^{\infty}[|(\mathbf{while})^{n}|](\rho_1))\\
&=\bigvee_{n=0}^{\infty}tr_{X-var(S)}([|(\mathbf{while})^{n}|](\rho_1))\\
&=\bigvee_{n=0}^{\infty}tr_{X-var(S)}([|(\mathbf{while})^{n}|](\rho_2))\\
&=tr_{X-var(S)}(\bigvee_{n=0}^{\infty}[|(\mathbf{while})^{n}|](\rho_2))\\
&=tr_{X-var(S)}([|\mathbf{while}|](\rho_2))
\end{split}\end{equation*} by continuity of trace.
\end{document}